\documentclass[
  aps,
  prper,
  reprint,
  superscriptaddress,
  amsmath,
  amssymb,
  letterpaper 
]{revtex4-2}

\usepackage{array}

\usepackage{graphicx}
\usepackage{dcolumn}
\usepackage{bm}
\usepackage{booktabs}
\usepackage[table]{xcolor}
\usepackage{url}
\usepackage{threeparttable}
\usepackage{float}
\usepackage{hyperref}

\begin{document}

\title{Navigating Quantum Education at the University-level: \\Pedagogical Pathways and a Roadmap for Textbook Selection}

\author{Yajie Liang}
\email{yajie.liang@uni-muenster.de}
\affiliation{Institute of Physics Education, University of Münster, Münster, Germany}
\affiliation{Department for Quantum Technology, University of Münster, Münster, Germany}

\author{Menno Poot}
\affiliation{Department for Quantum Technology, University of Münster, Münster, Germany}

\author{Carsten Schuck}
\affiliation{Department for Quantum Technology, University of Münster, Münster, Germany}
\affiliation{Center for Nanotechnology (CenTech), University of Münster, Münster, Germany}
\affiliation{Center for Soft Nanoscience (SoN), University of Münster, Münster, Germany}

\author{Stefan Heusler}
\affiliation{Institute of Physics Education, University of Münster, Münster, Germany}

\begin{abstract}
Instruction in quantum mechanics (QM) is often categorized by its entry point as “wave-first” or “spin-first,” yet these labels lack consistent definitions and overlook other established or emerging approaches. Moreover, while capturing broad differences in topic selection and sequencing, this binary framing obscures critical differences in contextualization—how three dimensions (mathematical tools, phenomena and experiments, and theoretical concepts) are framed and connected to support student sensemaking. 
Research also rarely examines how such pedagogical choices unfold across the undergraduate-to-graduate trajectory that shapes the contemporary landscape of university quantum education. Using textbooks as an analytical lens to map these choices and characterize overarching pathways—encompassing entry points and overall orientations—we comparatively analyze the selection, sequencing, and contextualization of core topics across these three dimensions, alongside prefaces and introductions, in 36 undergraduate- to graduate-level QM textbooks spanning countries and eras. At the topic level, we show that identical topics occupy widely different curricular roles depending on their placement and contextualization. Synthesizing these findings into broader pathways, we identify three additional entry-point archetypes—labeled path-, qubit-, and axiom-first—and characterize all five approaches, showing that most textbooks combine multiple archetypes concurrently or sequentially in their opening chapters. Extending this synthesis beyond entry points, we map textbooks’ overall orientations relative to mathematics, experiment and application, and theoretical physics. Interpreting these profiles alongside authors’ rationales and physics education research reveals how educators navigate competing trade-offs among mathematical accessibility, theoretical generality, physical motivation, and modern developments. This study provides a multidimensional framework, substantive examples, and a practical roadmap for textbook selection and curriculum design in quantum education.

\end{abstract}

\maketitle

\section{Introduction}
In quantum mechanics (QM) instruction, there is no unified consensus on teaching methodologies \cite{McKagen_2010, Kruse_2025, Griffiths_2018}. In particular, considerable debate in university-level instruction has centered on the distinction between curricula with ``wave-first'' (or ``position-first'') and with ``spin-first'' as entry point, which are strongly associated with instructors' textbook choices \cite{Buzzell_2025, Kruse_2025}.

Although the wave--spin distinction captures an important feature of QM instruction, this binary framing does not adequately represent contemporary pedagogical diversity. First, the labels lack consistent definitions: ``wave-first'' may denote a historical introduction via matter waves and wave--particle duality or a direct entry into continuous-variable wavefunctions and differential equations (often termed ``position-first''), whereas ``spin-first'' may refer narrowly to an introduction through the Stern--Gerlach experiment or more broadly to any approach beginning with a two-state system \cite{Pollock_2023, Riihiluoma_2025, Albert_2025}. 

Second, this dichotomy does not readily accommodate other established or emerging entry points---such as introductions based directly on abstract formalism and Dirac notation \cite{Levy_2025}, or approaches centered on sum-over-paths, quantum fields, or quantum technologies \cite{Taylor_1998, Hobson_2013, Giliberti_2024, carlini_2026}. While efforts to extend this vocabulary have identified and analyzed a wider range of approaches---spanning historical, matter-wave, two-state, path-integral, field-theoretic, and quantum-technology approaches---much of this broader classification work has focused on pre-university education \cite{Michelini_2021_2, Michelini_2021, Faletic_2024}. Although pre-university typologies provide a useful starting point, university-level instruction must additionally accommodate greater mathematical rigor, conceptual depth, and preparation for advanced study---while also shaping how future secondary teachers themselves conceptualize QM.

Third, although prior studies show that ``wave-first'' and ``spin-first'' curricula differ significantly in topic selection and sequencing \cite{Buzzell_2025, Kruse_2025}, this binary framing obscures critical differences in how topics are \textit{contextualized}. While the two labels are often associated with wave mechanics versus matrix representations or Dirac notation \cite{Riihiluoma_2025, Riihiluoma_2025_2, Levy_2025}, instructional differences extend well beyond the choice of mathematical formalism or initial physical example: curricula also vary in the relative emphasis placed on formal postulates versus experimental motivation \cite{Pollock_2023, Sadaghiani_2015}, and even the pedagogical value of two-state systems for introducing measurement depends heavily on the concrete physical context \cite{Bitzenbauer_2024}. These nuances matter directly for student learning. Research shows that students’ difficulties with QM arise not from mathematical complexity alone, but from the interplay of formalism, physical meaning, and interpretation \cite{Singh_2015, Lewerissa_2017}---including relating mathematical structures to physical phenomena, interpreting counterintuitive behavior, adopting probabilistic reasoning, and navigating the limitations of classical language \cite{Bouchée_2022}. Thus, beyond topic selection and sequencing, instructional approaches differ fundamentally in \textit{contextualization} across multiple dimensions: specifically, how mathematical tools, phenomena and experiments, and theoretical concepts are framed and connected to one another.

Beyond these limitations in characterizing how QM is initially introduced, research rarely examines how pedagogical choices---including the selection, sequencing and contextualization of topics---unfold and build upon one another across the broader undergraduate-to-graduate trajectory. At the tertiary level, physics education research (PER) has tended to separate introductory, upper-division, and graduate instruction rather than examining quantum education holistically \cite{Kruse_2025, Buzzell_2025, Robbins_2025}. This fragmented focus overlooks critical pedagogical transitions. As students advance to higher-level courses, they can often mask a lack of conceptual understanding behind adept mathematical calculation \cite{Singh_2015}. 
Navigating this trajectory is further complicated by students’ increasingly diverse academic preparation \cite{scherrer_2024, Dick_2020} and by contemporary curricular shifts. First, quantum information science and technology (QIST) has broadened the applications and instructional audiences of QM \cite{Doyle_2026, Meyer_2024}---with growing relevance in high-energy and condensed-matter physics \cite{Afik_2025, Zeng_Chen_Zhou_Wen_2019}---pushing historically advanced topics into earlier stages of instruction. Second, the growing availability of computational tools and large language models prompts reconsideration of how instruction across all levels should balance routine calculation with conceptual understanding, physical interpretation, and the critical evaluation of solutions. 

Taken together, these challenges underscore that we still lack a broad comparative picture of the patterns and variations in pedagogical choices that characterize the contemporary landscape of university quantum education. To map this landscape, textbooks provide an ideal analytical lens. Textbooks play a central role in education not only by shaping instructors’ curricular choices and guiding learners, but also as durable records that embody the pedagogical strategies, priorities, and instructional philosophies of their authors \cite{Vojír_2019, Ventura_2026}. Yet despite their importance, comparative research on university-level QM textbooks remains limited: while course-level surveys in the US document the continuing dominance of texts such as Griffiths \cite{Griffiths_2018} alongside the growing adoption of alternative textbooks \cite{Dubson_2009, Buzzell_2025, Kruse_2025}, much of the broader textbook literature either focuses on pre-university education \cite{Krijtenburg-Lewerissa_2019, Stadermann_2019, Merzel_2024_2, Winkler_2021} or examines only a small sample of university texts in relation to isolated concepts \cite{Emigh_2020, Mohan_2020, Zuccarini_2020}.

Against this background, this study develops and applies a framework built around three dimensions---mathematical tools, phenomena and experiments, and theoretical concepts---to examine how authors' topic-level \textit{pedagogical choices} shape both the granular organization of textbook content and the overarching \textit{pedagogical pathways}, alongside authors’ stated rationales. Within this framework, we characterize a textbook's pedagogical pathway through two complementary features across the three dimensions: its \textit{entry point} (the perspective through which QM is initially motivated and systematically developed) and its overall \textit{pedagogical orientation} (the broader priorities sustained throughout the text relative to the three dimensions). Treating each textbook as a holistic curricular design, we analyze the structure---or ``global topology'' \cite{Levy_2026}---of QM textbooks spanning undergraduate and graduate levels to address three research questions:

\begin{enumerate}
    \item[\textbf{RQ1:}] \textbf{Topic-level pedagogical choices:} What patterns and variations characterize these topic-level pedagogical choices---specifically, the selection, sequencing, and contextualization of core topics across the three dimensions in QM textbooks?
    \item[\textbf{RQ2:}] \textbf{Pedagogical pathways:} Synthesizing the findings from RQ1, how can textbooks' pedagogical pathways be characterized:
    \begin{enumerate}
        \item[\textbf{RQ2.1:}] With what concrete pedagogical choices can the familiar ``wave-first'' and ``spin-first'' entry-point labels be associated in textbooks?
        \item[\textbf{RQ2.2:}] What additional entry-point archetypes can be identified and characterized, and how do all the archetypes manifest in textbooks' opening chapters?
        \item[\textbf{RQ2.3:}] Beyond initial entry points, how can the overall pedagogical orientations of textbooks be characterized?
    \end{enumerate}
\end{enumerate}
To interpret the pedagogical reasoning underlying these structural patterns, we examine both RQs alongside authors’ stated rationales. 

Following an overview of our textbook corpus and analytical methods in Sec.~\ref{sec:methods}, our results progress from granular topic-level analyses to synthesized pathways. Section~\ref{sec:res_RQ1} examines topic-level pedagogical choices across the three dimensions (RQ1). Section~\ref{sec:res_RQ2} synthesizes these findings to characterize entry-point archetypes and overall pedagogical orientations (RQ2), culminating in a practical roadmap that translates the framework into guiding questions for textbook selection and curriculum design (Sec.~\ref{sec:res_4.3}). Finally, we discuss overall limitations and suggest future work in Sec.~\ref{sec:conclusion}.

\section{Methodology}
\label{sec:methods}

\subsection{Corpus Construction}
\label{sec:corpus_construction}

To address our research questions, we assembled a purposively selected corpus of university-level QM textbooks. To document the construction of this corpus transparently---from defining eligibility criteria (Sec.~\ref{sec:eligibility}) to identifying candidate texts and purposively selecting the final sample (Sec.~\ref{sec:selection})---we adapted selected reporting elements of the PRISMA guidelines \cite{Moher_2009}, using them strictly for reporting transparency rather than to imply an exhaustive systematic review.

\subsubsection{Eligibility criteria}
\label{sec:eligibility}
We considered English-language QM textbooks and published lecture notes mainly for undergraduate or graduate physics students, from the earliest publications in the field through September 2026. Where a text had multiple editions, we considered its latest edition. 

To support comparison of curricular organization across texts intended for general university QM courses in physics, we restricted the corpus to works offering a sufficiently broad treatment of core QM. Therefore, we excluded the following:
\begin{itemize}
    \item \textbf{Highly specialized formulations:} Texts devoted exclusively to a single mathematical formalism, such as path integrals \cite{kleinert_2009}, or matrix representation \cite{ludyk_2017}. 
    \item \textbf{Interpretational or historical works:} Books concerned primarily with the philosophy or foundations of QM \cite{durr_2020}, or with its historical development \cite{Longair_2013}.
    \item \textbf{Texts for other audiences:} Popular-science books \cite{Raymer_2017, Wu_2023}, treatments aimed primarily at engineers or chemists \cite{blinder_2020, Levi_2023}, or at mathematicians \cite{woit_2017, Givental_2019}, and supplementary study guides \cite{michelsen_2014, schwichtenberg_2018}. 
    \item \textbf{Other formats:} Problem collections, ``modern physics'' textbooks covering QM alongside a broad range of other topics \cite{krane_2019}, and highly condensed introductory texts \cite{Susskind_2014}.
    \item \textbf{Quantum-information-specific texts:} Books devoted exclusively to quantum information or quantum computing. \cite{Nielsen_Chuang_2010} Texts with a quantum-information emphasis remained eligible if they also offered a broad treatment of foundational QM suitable for the present curricular comparison; Haroche '06$^1$ was included on this basis. 
\end{itemize}

Two unpublished lecture notes were identified incidentally during corpus construction and would be excluded based on the publication-status criterion \cite{Styer_2024, Levy_2026}. Both otherwise met the eligibility criteria and contributed curricular perspectives not represented in the published texts selected for analysis, and therefore were still included. We did not undertake a dedicated search for unpublished lecture notes. 

\subsubsection{Identification of Eligible Texts and Study Selection}
\label{sec:selection}
Through targeted searches of academic publisher catalogs and reference tracing, we identified 140 QM texts that met the eligibility criteria specified above, including the two unpublished lecture notes admitted as exceptions. These texts formed the pool for subsequent purposive selection. Particularly rich reference sources included the extensive, chapter-level bibliographies in two texts \cite{Levy_2026, Cohen_2020}.

An examination of front matter across this initial pool revealed that the vast majority originated directly from university courses. These course origins spanned institutions across North America, Europe, Latin America, the Middle East, South Asia, and East Asia, ensuring that the candidate pool captured pedagogical traditions rooted in diverse national educational systems—many of which have attained international reach through translations.

From the candidate pool, one author purposively selected texts for detailed analysis, balancing established works with recent publications. Rather than aiming for an exhaustive or statistically representative sample, this selection sought to maximize variation in curricular organization to support comparison across diverse instructional approaches. Consequently, otherwise eligible candidate texts were excluded when their organization closely duplicated approaches already captured in the corpus; for instance, several texts \cite{zettili_2022, reed_2022, wohl_2022, scherrer_2024, Cole_2024} were omitted because their curricular frameworks closely mirrored the traditional wave-mechanics trajectory exemplified by Schiff \cite{schiff_1968}.

Table~\ref{table:books} lists the 36 texts in the resulting corpus along with their examined editions, bibliographic references, and compact text IDs of the form [Author 'YY$^{\text{Ed}}$] (e.g., [Griffiths '18$^3$]; superscripts are omitted for texts not published), which we use to refer to corpus texts throughout the remainder of the paper. Moreover, for each of the 36 selected texts, we compiled descriptive metadata to contextualize the instructional focus of each text and documented it in Table~\ref{table:books}:
\begin{itemize}
    \item \textbf{Intended audience and institutional course origins:} Extracted directly from explicit statements in prefaces or forewords.
    \item \textbf{Author metadata at the time of publication:} Institutional affiliations and primary research specializations were gathered from author biographies and verified against institutional directories and scholarly databases where necessary.
\end{itemize}
These contextual attributes were used for corpus presentation and interpretation, rather than as additional outcome categories.

To contextualize comparisons of curricular organization, we distinguish texts by their stated prerequisites. Throughout this paper, \emph{introductory texts} refer to those that assume no prior knowledge of QM and \emph{advanced texts} to those that do---a distinction based solely on prior QM exposure rather than mathematical prerequisites or the level of topics covered.

\subsection{Data Collection, Codebook Development, and Pathway Synthesis}
\label{sec:data_collection}
To complement the structural analysis across both RQ1 and RQ2, one author conducted a qualitative review of textbook prefaces and introductory chapters, extracting and synthesizing authors' explicit statements regarding target audiences, prerequisite assumptions, instructional philosophies, and rationales for content selection, sequencing, and contextualization.

\subsubsection{Topic-Level Coding and Analysis (RQ1)}
To address RQ1 concerning topic-level pedagogical choices---and to provide the empirical foundation for characterizing overarching pathways (RQ2)---we selected core topics within each of the three dimensions (mathematical tools, phenomena and experiments, and theoretical concepts) motivated by prior PER and textbook literature on student sensemaking, curricular transitions, and disciplinary scope: (i) \textit{mathematical formalisms} (wave mechanics, matrix mechanics, Dirac notation, path integrals, density operators, and second quantization); (ii) \textit{initial physical systems} (first worked infinite-dimensional and finite-dimensional systems, qubits, and first conceptual single-photon systems); and (iii) \textit{theoretical concepts}, spanning foundational postulates (state, observable, measurement outcome, measurement probability, post-measurement state, time evolution, composite systems, and symmetrization), physical principles (superposition, complementarity, uncertainty, correspondence, Pauli’s exclusion principle, and the no-cloning theorem), and topics in foundations and interpretations (entanglement, nonlocality, interpretations of QM, the measurement problem, and decoherence). To clarify how these topic areas inform our RQs and to keep the empirical findings closely tied to their pedagogical context, we outline the specific PER and textbook background motivating each set of topics at the start of its corresponding subsection in Sec.~\ref{sec:res_RQ1} before reporting the corpus data.

To systematically record the \textit{selection} and \textit{sequencing} of these core topics, we developed a structured codebook (provided in the Supplementary Material \ref{sec:codebook}). Candidate occurrences of each topic were first located using tables of contents, indices, and full-text searches, and then verified through close reading of the surrounding text. Within-topic distinctions and operational rules for recording topic placement were iteratively refined during coding, with earlier entries revisited whenever coding rules were updated. To establish coding reliability and clarity, a second coder independently applied the provisional codebook to four randomly selected textbooks; discrepancies were discussed to refine the coding definitions and reach full consensus.

To analyze \textit{contextualization}, we qualitatively examined the instructional roles and framing of each selected topic within its surrounding chapter and section structure. Specifically, we documented how topics were embedded across the three dimensions as well as how they were connected to other topics. For topics appearing multiple times within a book, we also tracked how their framing evolved across chapters. Cross-text comparisons were then used to identify recurring contextualization patterns and notable departures, which are illustrated through representative examples in Sec.~\ref{sec:res_RQ1}.

\subsubsection{Synthesis of Pedagogical Pathways (RQ2)}
To address RQ2, we synthesized our topic-level findings (RQ1) into a two-part analytical framework (Sec.\ref{sec:res_RQ2}): \textit{entry-point archetypes} (Table\ref{table:qm_pedagogy}; RQ2.1--RQ2.2) and \textit{pedagogical orientation criteria} across the three dimensions (Table~\ref{table:criteria}; RQ2.3). To demonstrate the framework's utility, one author applied these tools to construct an indicative profile for each of the 36 textbooks (Sec.~\ref{sec:res_4.3}).

Specifically, entry points were coded from each textbook's initial development of QM, excluding purely preparatory material, such as mathematical tools presented without explicit connection to QM. The scope was determined by this introductory development rather than a fixed number of chapters, capturing both single-archetype openings and multi-archetype combinations.
Overall orientations were rated against Table~\ref{table:criteria} using authors' rationales as complementary evidence. Finally, to provide additional context, the same author reviewed tables of contents to record brief descriptors of disciplinary scope and specialized thematic emphasis (leaving general texts unlabeled).

\begin{table*}[h]
\caption{Textbook corpus with the examined versions, copyright year, and descriptive metadata. Textbook IDs identify the texts examined and are used throughout the analysis; numbered citations link to the full bibliographic records.}
\label{table:books}
\footnotesize
\renewcommand{\arraystretch}{1.2}
\resizebox{1.0\textwidth}{!}{\begin{tabular}{|p{0.18\textwidth}|p{0.26\textwidth}|p{0.08\textwidth}|p{0.10\textwidth} |p{0.22\textwidth}|p{0.16\textwidth}|}

\toprule
\textbf{Text ID} &
\textbf{Title} &
\textbf{Ed. / year} &
\textbf{Intended audience} &
\textbf{Affiliation(s)} &
\textbf{Research field(s)} \\
\midrule

\multicolumn{6}{l}{\textbf{Reference and comprehensive texts (9)}}\\
\addlinespace
\textbf{Feynman '11 ('65)} \cite{Feynman_2011} &
The Feynman Lectures on Physics, Vol.~3 &
1965 &
Introductory &
Caltech (US) &
Theory \\

Dirac '67$^4$ \cite{Dirac_1967} &
The Principles of QM &
4th / 1967 &
 &
University of Cambridge$^*$ (UK) &
Theory \\

Shankar '94$^2$ \cite{Shankar_1994} &
Principles of QM &
2nd / 1994 &
All levels &
Yale University (US) &
Theory \\

Messiah '95$^2$ \cite{Messiah_1995} &
QM &
2nd / 1995 &
All levels &
Center of Nuclear Studies (France) &
Theory \\

Auletta '09$^1$ \cite{Auletta_Fortunato_Parisi_2009} &
QM &
1st / 2009 &
All levels; all backgrounds &
Multiple universities in Rome$^*$ (Italy) &
Philosophy; financial engineering; theory \\

Cohen-T. '20$^2$ \cite{Cohen_2020} &
QM, 3 vols. &
2nd / 2020 &
UG; G &
Multiple universities$^*$ (France) &
Experiment; theory; experiment \\

Zwiebach '22$^1$ \cite{Zwiebach_2022} &
Mastering QM &
1st / 2022 &
UG; G &
MIT (US) &
String theory \\

Cetto '26$^1$ \cite{Cetto_2026} &
QM: A Physical Approach &
1st / 2026 &
UG; G; all backgrounds &
National Autonomous University of Mexico$^*$ (Mexico) &
Theory; theory \\

Freericks '26$^1$ \cite{Freericks_2026} &
QM Done Right: The Shortest Path from Novice
to Researcher &
1st / 2026 &
All levels &
Georgetown University$^*$ (US) &
Theory \\
\midrule

\multicolumn{6}{l}{\textbf{Undergraduate texts (12)}}\\
\addlinespace
Liboff '03$^4$ \cite{Liboff_2003} &
Introductory QM &
4th / 2003 &
Introductory &
University of Central Florida$^*$ (US) &
Theory \\

Binney '14$^1$ \cite{Binney_2014} &
The Physics of QM &
1st / 2014 &
Introductory &
University of Oxford (UK) &
Theoretical astrophysics; HEP theory \\

Dunningham '18$^2$ \cite{Dunningham_2018} &
Introductory Quantum Physics and Relativity &
2nd / 2018 &
Introductory &
Universities (UK) &
Theory; QI theory \\

Griffiths '18$^3$ \cite{Griffiths_2018} &
Introduction to QM &
3rd / 2018 &
UG &
Reed College$^*$ (US) &
Theory; CM theory \\

\textbf{Lvovsky '18$^1$} \cite{Lvovsky_2018} &
Quantum Physics: An Introduction Based on Photons &
1st / 2018 &
UG &
University of Calgary (Canada) &
Experimental quantum optics \\

\textbf{Pade '18$^2$} \cite{Pade_2018_1,Pade_2018_2} &
QM for Pedestrians: Fundamentals; \newline
QM for Pedestrians 2: Applications and Extensions &
2nd / 2018 \newline  1st / 2014 &
Introductory &
University of Oldenburg (Germany) &
Theory \\

Banks '19$^1$ \cite{Banks_2019} &
QM: An Introduction &
1st / 2019 &
UG; beginning G &
Rutgers University$^*$ (US) &
HEP theory \\

McIntyre '22$^1$ \cite{Mcintyre_2022} &
QM: A Paradigms Approach &
1st / 2022 &
UG &
Oregon (US) &
Experimental optics \\

\textbf{Basdevant '23$^3$} \cite{Basdevant_2023} &
Lectures on QM &
3rd / 2023 &
Introductory &
\'Ecole Polytechnique (France) &
Theory \\

\textbf{Kok '23$^2$} \cite{Kok_2023} &
A First Introduction to Quantum Physics &
2nd / 2023 &
Introductory &
University of Sheffield$^*$ (UK) &
Optics; QI theory \\

Larkoski '23$^1$ \cite{Larkoski_2023} &
QM: A Mathematical Introduction &
1st / 2023 &
Introductory &
SLAC National Accelerator Laboratory (US) &
Particle theory \\

\textbf{Styer '24} \cite{Styer_2024} &
The Physics of QM &
2024 &
Introductory &
Oberlin College (US) &
Theory \\
\midrule

\end{tabular}}
\end{table*}


\begin{table*}[t]
\addtocounter{table}{-1} 
\caption{Textbook corpus with the examined versions, copyright year, and descriptive metadata. Textbook IDs identify the texts examined and are used throughout the analysis; numbered citations link to the full bibliographic records. (Continued)}
\label{table:books_part2}
\begin{threeparttable}
\footnotesize
\renewcommand{\arraystretch}{1.2}
\resizebox{1.0\textwidth}{!}{\begin{tabular}{|p{0.18\textwidth}|p{0.26\textwidth}|p{0.08\textwidth}|p{0.10\textwidth}|p{0.22\textwidth}|p{0.16\textwidth}|}
\toprule
\textbf{Text ID} & \textbf{Title} & \textbf{Ed./year} & \textbf{Intended audience} & \textbf{Affiliation(s)} & \textbf{Research field(s)} \\
\midrule

\multicolumn{6}{l}{\textbf{Advanced-undergraduate and beginning-graduate texts (5)}}\\
\addlinespace
\cellcolor{gray!25}Haroche '06$^1$ \cite{Haroche_2006} &
Exploring the Quantum: Atoms, Cavities, and Photons &
1st / 2006 &
Advanced UG &
Diverse universities (France) &
Experiment; experiment \\

\cellcolor{gray!25}Beck '12$^1$ \cite{Beck_2012} &
QM: Theory and Experiment &
1st / 2012 &
Advanced UG &
Whitman College$^*$ (US) &
Experiment \\

\cellcolor{gray!25}Townsend '12$^2$ \cite{Townsend_2012} &
A Modern Approach to QM &
2nd / 2012 &
Advanced UG &
Harvey Mudd College (US) &
Particle theory \\

\cellcolor{gray!25}Dick '20$^3$ \cite{Dick_2020} &
Advanced QM: Materials and Photons &
3rd / 2020 &
Advanced UG; G &
University of Saskatchewan (Canada) &
Theory \\

\cellcolor{gray!25}Simon '25$^1$ \cite{Simon_2025} &
Introduction to Quantum Science and Technology &
1st / 2025 &
Advanced UG; G; all backgrounds &
Stonehill College (US) &
Experiment \\
\midrule

\multicolumn{6}{l}{\textbf{Graduate texts (10)}}\\
\addlinespace
\cellcolor{gray!45}Landau '77$^3$ \cite{Landau_1977} &
QM: Non-Relativistic Theory
(Course of Theoretical Physics, Vol.~3) &
3rd / 1977 &
G &
(Soviet Union) &
Theory; theory \\

\cellcolor{gray!45}\textbf{Jackson '24 ('89)} \cite{Jackson_2024} &
A Course in QM &
1st / 2024 &
G &
University of California (US) &
HEP theory \\

\cellcolor{gray!45}Gottfried '03$^2$ \cite{Gottfried_2003} &
QM: Fundamentals &
2nd / 2003 &
G &
Cornell University$^*$ (US) &
Theory; particle theory \\

\cellcolor{gray!45}Commins '14$^1$ \cite{Commins_2014} &
QM: An Experimentalist's Approach &
1st / 2014 &
G &
University of California, Berkeley (US) &
Experiment \\

\cellcolor{gray!45}Ballentine '15$^2$ \cite{Ballentine_2015} &
QM: A Modern Development &
2nd / 2015 &
G &
Simon Fraser University (Canada) &
Theory \\

\cellcolor{gray!45}Wan '19$^1$ \cite{Wan_2019} &
QM: A Fundamental Approach &
1st / 2019 &
G &
University of St Andrews (UK) &
Theory \\

\cellcolor{gray!45}Sakurai '21$^3$ \cite{Sakurai_2021} &
Modern QM &
3rd / 2021 &
G &
(US) &
Particle theory; nuclear experiment \\

\cellcolor{gray!45}Năstase '23$^1$ \cite{Nastase_2022} &
QM: A Graduate Course &
1st / 2023 &
G &
(Brazil; US) &
String theory \\

\cellcolor{gray!45}\textbf{Levy '26} \cite{Levy_2026} &
From Qubit to Qubit &
2026 &
G &
Pittsburgh (US) &
CM experiment \\

\cellcolor{gray!45}Martin '26$^1$ \cite{Martin_2026} &
QM: A Comprehensive Graduate Text &
1st / 2026 &
G &
Northern Illinois University$^*$ (US) &
HEP theory \\

\bottomrule

\end{tabular}
}

 \begin{tablenotes}
\item \textit{Abbreviations:} CM = condensed matter physics; G = graduate; HEP = high-energy physics; QI = quantum information; QM = quantum mechanics; UG = undergraduate. Research fields are listed in author order.
\item \textit{Note}: An asterisk ($\ast$) indicates that a university-course origin could not be established from the consulted sources.
\item Texts are grouped by scope and target audience and ordered chronologically by copyright year within each group—except for Feynman '11 ('65) and Jackson '24 ('89), which are ordered according to the dates of their underlying lectures (1961--1964, and 1978--1979 / 1988--1989, respectively).
\item Across all tables, unshaded (white) and gray-shaded ID cells distinguish texts that assume no prior QM knowledge (\textit{introductory}) from those that do (\textit{advanced}), respectively.
\item \textbf{Bold Text IDs} denote lecture-based texts (Feynman '11 ('65), Basdevant '23$^3$, Jackson '24 ('89)), volumes in Springer's \textit{Undergraduate Lecture Notes in Physics} series p(Lvovsky '18$^1$, Pade '18$^2$, Kok '23$^2$), or unpublished lecture notes (Styer '24, Levy '26).
 \end{tablenotes}
 
\end{threeparttable}

\end{table*}

\section{Topic-level pedagogical choices and authors' rationales}
\label{sec:res_RQ1}
We begin by examining authors' topic-level pedagogical choices---the selection, sequencing, and contextualization of core topics (RQ1) across the three dimensions of our framework: \textit{mathematical tools} (Sec.~\ref{sec:res_1}), \textit{phenomena and experiments} (Sec.~\ref{sec:res_2}), and \textit{theoretical concepts} (Sec.~\ref{sec:res_3}). Each subsection first outlines the pedagogical background motivating the selected core topics and their connection to our RQs, before synthesizing empirical patterns across our 36-textbook corpus (Tables~\ref{table:1}--\ref{table:3.3}) with authors' stated rationales---thereby integrating empirical results and pedagogical discussion directly within each section rather than in a separate discussion section. To keep the main text focused on overarching structural patterns, cross-dimensional connections, and pedagogical trade-offs, detailed textbook-by-textbook breakdowns, illustrative examples, and specific citations for authors' rationales are provided in the Supplementary Material.

\subsection{Mathematical Tools: Mathematical formalisms}
\label{sec:res_1}

In physics, mathematics serves both computational and conceptual functions: it represents physical ideas, structures reasoning, and enables new insights \cite{Uhden_2012}. This role is especially important in QM, which can be expressed through numerous mathematical formulations, including differential equations, linear operators, group-theoretic methods, and functional analysis \cite{Styer_2002}.

Given this diversity, our topic-level analysis (RQ1) focuses on five widely taught formalisms in general QM courses: \textbf{wave mechanics, matrix mechanics, path integrals, density matrices, and second quantization}. We also treat \textbf{Dirac's bra--ket notation} as a separate analytical category. Although bra--ket notation is not itself a formulation of QM, it is closely tied to the abstract Hilbert-space framework and provides a basis-independent language connecting wave, matrix, and operator representations \cite{Gieres_2000}. Together with matrix representations, it also forms the principal mathematical language of QIST.

Prior literature notes that the initial sequencing of elementary representations—central to characterizing ``wave-first'' and ``spin-first'' entry points (RQ1, RQ2.1)—shapes students' understanding and subsequent preferences \cite{Riihiluoma_2025_2} \cite{Schermerhorn_2019}. These findings highlight competing pedagogical reasonings in the broader textbook literature:  wave-first sequencing is often justified through students' familiarity with differential equations and its clearer connection to classical physics \cite{Gasiorowicz_2003}, whereas abstract or spin-first approaches aim to foreground distinctively quantum concepts and avoid identifying QM exclusively with wave mechanics \cite{schwindt_2016}. Yet the relationship between formalism and notation is not absolute; Weinberg \cite{Weinberg_2015}, for example, adopts an abstract Hilbert-space framework conceptually aligned with Dirac, yet deliberately avoids using Dirac notation.

Beyond initial sequencing, how textbooks contextualize and connect multiple representations (RQ1) is critical for addressing documented learning difficulties. Fluency across representations is an important objective of QM instruction because different problems and disciplinary traditions privilege different mathematical tools \cite{Emigh_2020, Riihiluoma_2025}. Nevertheless, students often struggle to interpret and translate between wavefunctions, state vectors, matrices, and Dirac notation \cite{Wawro_2020, Gire_2015}. For example, conflating a wavefunction with an abstract state vector can impede probabilistic reasoning, even among advanced students \cite{Wan_Emigh_2019, Singh_2015}.

Finally, beyond elementary representations, the selection, sequencing, and contextualization of path integrals, density operators, and second quantization offer further insight into a textbook's overall pedagogical orientation across the undergraduate-to-graduate trajectory (RQ1, RQ2.3). Path integrals provide an alternative formulation of quantum dynamics through propagators and action-based amplitudes, connecting QM to statistical mechanics and field theory. Density operators generalize the description of quantum states to include mixed states and reduced states of subsystems, providing tools for the treatment of open systems and experimental preparation and measurement. Second quantization provides a framework for describing identical particles and quantized fields, connecting QM to many-body physics, quantum optics, and quantum field theory. 

Table~\ref{table:1} summarizes the selection and sequencing of these six mathematical formalisms across our 36-textbook corpus, which we examine below across initial representations, other formalisms, and strategies for connecting representations.

\begin{table*}[h!]
\caption{Chapter-level placement of the first introduction of selected mathematical formalisms in QM texts. Blue shading identifies the earliest introduced formalism(s), including cases in which multiple formalisms are introduced within the same chapter.
Green shading identifies cases in which at least 3 formalisms are introduced together in a subsequent chapter following an initial formalism.\\
}
\label{table:1}
\begin{threeparttable}
\centering
\footnotesize
\renewcommand{\arraystretch}{1.1}
\resizebox{\textwidth}{!}{
\begin{tabular}{|l|c|c|c|c|c|c|}
\toprule
{Text ID} & \textbf{\shortstack{Wave\\mechanics}} & \textbf{\shortstack{Matrix\\mechanics}} & \textbf{\shortstack{Dirac\\notation}} & \textbf{\shortstack{Path-\\integral}} & \textbf{\shortstack{Density\\matrix}} & \textbf{\shortstack{Second\\quantization}}   \\ 
\hline
\midrule

\textbf{Feynman '11 ('65)} & 16/21 & \cellcolor{blue!25}8/21 & \cellcolor{blue!25}8/21 & / & / & /  \\ 
Dirac '67$^4$ & 3/12 & (3)/12 & \cellcolor{blue!25}1/12 & / & / & 10/12   \\ 
Shankar '94$^2$ & (4)/21 & \cellcolor{blue!25}(1)/21 & \cellcolor{blue!25}1/21 & 8, 21/21  & 4/21 & /  \\ 
Messiah '95$^2$ & \cellcolor{blue!25}2/21 & 2$^*$, (8)/21 & 7/21 & / & 8/21 &  (21)/21  \\  
Auletta '09$^1$ & 2/17 & \cellcolor{blue!25}(1), 2/17 & 
\cellcolor{blue!25}1/17 & 10/17 & 5, 9, 14/17 & 13/17  \\ 
Cohen-T. '20$^2$   & \cellcolor{blue!25}(1)/21 & (2)/21 & 2/21 & App. 4 & 3/21 & 15/21 \\ 
Zwiebach '22$^1$ & (3)/30 & \cellcolor{blue!25}(2)/30 & 12/30 & / & 22/30 & /  \\ 
Cetto '26$^1$ &  4/19 & \cellcolor{blue!25}3/19  & 4/19 & / &  15/19 & 17/19  \\ 
Freericks '26$^1$ & \cellcolor{blue!25}5.8, 11/23 & \cellcolor{blue!25}(5.7)/23 & \cellcolor{blue!25}5.8/23 & / & \cellcolor{blue!25}5.8, 22/23 & 17/23  \\ 
\midrule

Liboff '03$^4$ & \cellcolor{blue!25}(2)/16 & 11/16 & 4/16 & 7/16 & 11/16 & 7, 13/16 (ex.) \\ 
Binney '14$^1$ & (2)/13 & (2)/13 & \cellcolor{blue!25}1/13 &  / &  6/13 & / \\ 
Dunningham '18$^2$ & \cellcolor{blue!25}3/13 & 8/13 & 8/13 &  / & 8/13 & (8), (11)/13 \\ 
Griffiths '18$^3$ & \cellcolor{blue!25}(1)/12 & (3), (4)/12 & 3/12 & / & 12/12 & /  \\ 
\textbf{Lvovsky '18$^1$} &  3/5 & \cellcolor{blue!25}(1)/5 & \cellcolor{blue!25}1/5 & / & 5/5 & 3$^*$ /5 \\ 
\textbf{Pade '18$^2$}  & \cellcolor{blue!25}1, 11/28 & (2), 11/28 & 2, 11/28 & / & 22/28 &  App. W \\ 
Banks '19$^1$ & (3)/21 & \cellcolor{blue!25}(2)/21 & \cellcolor{blue!25}2, 6/21 & 19/21 & \cellcolor{blue!25}2/21 & (5), (12)/21  \\ 
McIntyre '22$^1$ & (5)/16 & \cellcolor{blue!25}(1)/16 & \cellcolor{blue!25}1/16 & / & / & (9)$^*$/16 \\ 
\textbf{Basdevant '23$^3$} & \cellcolor{blue!25}3/19 & 7/19 & 6/19 & /  & / & /  \\ 
\textbf{Kok '23$^2$} & 8/11 & \cellcolor{blue!25}(2), 8/11 & \cellcolor{blue!25}2/11 & / &  7/11  & /  \\ 
Larkoski '23$^1$ &  \cellcolor{blue!25}(3.5)/13 & \cellcolor{blue!25}(3.4)/13 & \cellcolor{blue!25}3.4/13  & 11/13 & 4, 12/13  &  / \\ 
\textbf{Styer '24} &  6/24 & \cellcolor{blue!25}(2)/24 & \cellcolor{blue!25}2/24 &  / & 4$^{*}$/24 & / \\ 
\midrule

\cellcolor{gray!15} Haroche '06$^1$ & \cellcolor{blue!25}(2.1)/10 & \cellcolor{blue!25}(2.1/10)  & \cellcolor{blue!25}2.1/10 & / &  \cellcolor{blue!25}2.4/10 &  9/10 \\ 
\cellcolor{gray!15} Beck '12$^1$ & 11/17 & \cellcolor{blue!25}(3)/17 & \cellcolor{blue!25}3/17 & / & / & (16)/17 \\ 
\cellcolor{gray!15} Townsend '12$^2$ & 6/14 & 2/14 & \cellcolor{blue!25}1/14 & 8/14 & 5/14 & 14/14  \\ 
\cellcolor{gray!15} Dick '20$^3$  & \cellcolor{blue!25}1/23 & (7)/23 & 4/23 & 14/23 & / & 17/23  \\ 
\cellcolor{gray!15} Simon '25$^1$  & \cellcolor{blue!25}(1)/27 & (3)/27 & \cellcolor{blue!25}1/27 & / & 3/27 & (5)/27 \\ 
\midrule

\cellcolor{gray!45} Landau '77$^3$ & \cellcolor{blue!25}1/18 & \cellcolor{green!25} (2)/18 & \cellcolor{green!25} 2/18  & / &  \cellcolor{green!25} 2/18 & 9/18  \\ 
\cellcolor{gray!45} \textbf{Jackson '24 ('89)} & \cellcolor{blue!25}1/10 & 
\cellcolor{green!25} (2)/10 & \cellcolor{green!25} 2/10 & / & \cellcolor{green!25} 2/10 & (8)/10 \\ 
\cellcolor{gray!45} Gottfried '03$^2$ & \cellcolor{blue!25}1/13 & (3)/13  & \cellcolor{green!25} 2/13 &  \cellcolor{green!25} 2/13 & \cellcolor{green!25} 2/13 & (10), 11/13  \\ 
\cellcolor{gray!45} Commins '14$^1$ & 4/25 & (3)/25 & \cellcolor{blue!25}2/25 & 6/25 & 3/25 & 11, (15)/25   \\ 
\cellcolor{gray!45} Ballentine '15$^2$ & (4)/21 & (2)/21 & \cellcolor{blue!25}1/21 & 4/21 & 2/21 & (17)/21  \\ 
\cellcolor{gray!45} Wan '19$^1$ & \cellcolor{blue!25}(10)/37  & (13)/37 & 16/37 & / & 18/37 &  (33)/37 \\ 
\cellcolor{gray!45} Sakurai '21$^3$ & \cellcolor{blue!25}1.7/8 & \cellcolor{blue!25}(1.3), 1.6/8 & \cellcolor{blue!25}1.2/8 & 2/8 & 3/8 & 7/8  \\ 
\cellcolor{gray!45} Năstase '23$^1$ &  (0$^{*}$), (3)/58 & 0$^{*}$, (4)/58 & \cellcolor{blue!25}1/58 & 10, 28/58 & 3, 32/58 & 56/58  \\ 
\cellcolor{gray!45} \textbf{Levy '26} & 3/12 & \cellcolor{blue!25}(1.10)/12 & \cellcolor{blue!25}1.3/12 & 12/12 & \cellcolor{blue!25}1.7/12 & 3/12  \\ 
\cellcolor{gray!45} Martin '26$^1$ & \cellcolor{blue!25}(2.8)/29 & \cellcolor{blue!25}(2.7)/29  & \cellcolor{blue!25}2.1/29  & 29/29  & 3/29 &  / \\ 
\bottomrule

\end{tabular}
}
 \begin{tablenotes}
 \item \textit{Note:} Parentheses \textbf{(~)} indicate discussion without explicit use of the term;
 \item an asterisk ($\ast$) indicates a relatively brief introduction.
\item \textit{Abbreviations:} ex.\ = exercise; App.\ = appendix.
 \end{tablenotes}
 \end{threeparttable}
\end{table*}

\subsubsection{Initial mathematical representations and entry points}

\paragraph{Dirac notation and the Schrödinger equation.}
Across our corpus, the topic-level data show that while some mathematical features commonly associated with ``wave-first'' and ``spin-first'' recur, these associations are not rigid. For instance, Dirac notation is frequently introduced alongside matrix representations in texts commonly labelled as ``spin-first,'' but other texts in the corpus separate their introduction: some use Dirac notation early to emphasize abstract states and measurement before developing explicit matrix calculations, whereas others introduce it later to unify wave and matrix mechanics.

Similarly, the initial contextualization of the Schrödinger equation generally reflects the mathematical language developed beforehand. It typically appears in position-space form in texts that open with wave mechanics, consistent with characterizations of ``wave-first'' instruction in the PER literature \cite{Buzzell_2025}, and appears in abstract form in texts that initially emphasize abstract states. Nevertheless, the initial mathematical language does not uniquely determine the subsequent treatment of dynamics. For example, Freericks '26$^1$ introduces several mathematical tools early and first presents the Schrödinger equation in Dirac notation, yet explicitly distinguishes his representation-independent ``operator mechanics'' from both conventional spin-first and ``Schrödinger-equation-first'' approaches \cite{Freericks_2023}.

\paragraph{Authors' rationales for initial mathematical sequencing.}
Authors motivate their initial mathematical sequencing through several recurring pedagogical considerations: reducing computational overhead, establishing representational generality, differentiating successive levels of instruction, and scaffolding abstraction. Some authors favor early algebraic or operator-based approaches to prevent differential-equation techniques from overshadowing conceptual development and to present wave mechanics as just one representation of a broader theory. Others emphasize the cognitive challenges of early abstraction, motivating either a gradual progression from familiar mathematical structures or a deliberate interweaving of analytical and algebraic approaches.

\subsubsection{Other mathematical formalisms}
Across the corpus, path integrals, density operators, and second quantization are not uniformly reserved for advanced or graduate instruction: some texts incorporate them into the initial development of the theory, whereas others present them as later extensions.

\paragraph{Path integrals.}
Path-integral coverage in the corpus ranges from omission to substantial early treatment. In their prefaces, authors' stated rationales balance conceptual insight and relevance to subsequent study against prerequisite mathematical knowledge and curricular time constraints.

\paragraph{Density operators.}
Density operators likewise occupy diverse pedagogical roles across the analyzed texts: they may accompany early matrix representations, form part of the general postulates, or arise later in discussions of mixed states, decoherence, and quantum information. Notably, late placement does not necessarily imply that authors regard them as peripheral: Townsend '12$^2$ explicitly emphasizes their central experimental relevance despite deferring their introduction.

\paragraph{Second quantization.}
Second quantization appears in almost all advanced texts, generally relatively late, but is sequenced earlier in some contemporary texts associated with QIST. Its contextualization commonly proceeds either through identical particles and many-body states or through quantization of the radiation field, connecting the formalism to different physical applications and disciplinary priorities. Authors' rationales for including second quantization emphasize its conceptual role in understanding wave--particle duality, particle identity, and quantum statistics; its mathematical accessibility via harmonic-oscillator methods; and its relevance to modern applications and preparation for quantum field theory (QFT).

\subsubsection{Connecting mathematical representations}
Concerning how textbooks contextualize different representations in relation to one another—sometimes including path integrals, density operators, or second quantization alongside elementary ones-our analysis identifies four primary organizational strategies for bringing them into contact: (i) introducing multiple representations through a shared physical setting; (ii) collecting mathematical tools in a dedicated interlude; (iii) integrating them within a unified general formal framework; or (iv) alternating analytical and algebraic approaches across successive chapters. At more advanced levels, Levy '26 explicitly connects particle-based (first-quantized) and field-based (second-quantized) descriptions.

\subsection{Phenomena and Experiments: First Physical Systems}
\label{sec:res_2}
Concrete physical systems and experimental contexts play a central role in QM instruction by grounding abstract formalism and shaping how students make sense of quantum concepts \cite{Sadaghiani_2015, Bitzenbauer_2024}; textbook authors such as Gasiorowicz \cite{Gasiorowicz_2003} also emphasize the role of simple physical models in developing physical intuition. 

Closely tied to the wave-first/spin-first debate is whether instruction begins with continuous, \textbf{infinite-dimensional} models or discrete, \textbf{finite-dimensional} systems (where dimensionality refers to the state space of the pedagogical model)---a choice that shapes both the mathematical tools required and the physical phenomena foregrounded (informing RQ2.1, RQ2.2). Infinite-dimensional models (such as potential wells) traditionally support wave-mechanical treatments centered on spatial dynamics, continuous probability distributions, and boundary-value problems. Conversely, finite-dimensional models (most notably two-state systems such as spin-$1/2$) pair naturally with matrix and operator mechanics, foregrounding distinctively quantum phenomena such as superposition and projective measurement.

To examine how textbooks navigate this choice, we first identify and compare the initial \textit{worked examples} formulated in {infinite-dimensional} and {finite-dimensional} Hilbert spaces. Beyond these, we also track two physical-system categories that reflect contemporary shifts (informing RQ2.3): the introduction of \textbf{qubits} and conceptual illustrations using \textbf{single-photon systems}. In our analysis, we distinguish \textit{worked examples}---physical systems subjected to step-by-step mathematical calculation or derivation yielding physically interpreted results (such as eigenstates, measurement probabilities, expectation values, or time evolution)---from \textit{conceptual illustrations}, which use a physical system primarily to motivate or interpret a quantum idea without sustained mathematical treatment. 

Two-state systems connect foundational QM instruction to modern developments in QIST through the qubit. Whereas specialized quantum-information texts---most notably Nielsen and Chuang \cite{Nielsen_Chuang_2010}---begin directly with qubits, general QM textbooks differ markedly in whether, when, and in what context they adopt this information-theoretic framing. Referring to abstract qubits rather than solely to specific physical realizations can also provide a shared conceptual language for learners from diverse disciplinary backgrounds \cite{Doyle_2026}. We therefore track the explicit introduction and contextualization of the \textit{qubit} as an indicator of engagement with QIST.

Similarly, we examine whether and how single-photon systems are used for conceptual illustrations beyond their traditional historical role in motivating energy quantization and wave--particle duality. Single-photon and quantum-optical experiments have become increasingly prominent in contemporary quantum education due to their conceptual transparency, relevance to quantum technologies, and accessibility in undergraduate laboratories \cite{Galvez_2005, Lukishova_2022, Borish_2023, Borish_2024, Doering_2025}.

Table~\ref{table:2} summarizes the selection and sequencing of these four physical-system categories across our 36-textbook corpus, which we examine below in terms of their alignment with mathematical sequencing and the specific systems chosen within each category.

\begin{table*}[h!]
\caption{Chapter-level placement of the first illustrative physical
systems in the selected QM texts. Blue shading identifies the earliest
illustrative system(s), whether used as worked examples or conceptual
illustrations; the introduction of the term \textit{qubit} is also highlighted when it occurs in the same chapter.}
\label{table:2}
\begin{threeparttable}
\centering
\footnotesize
\renewcommand{\arraystretch}{1.1}
\resizebox{\textwidth}{!}{\begin{tabular}{|l|l|c|l|c|c|l|c|}
\toprule
{Text ID} &  \multicolumn{2}{c|}{\textbf{\shortstack{First worked\\inf.-dim. system}}}  & \multicolumn{2}{c|}{\textbf{\shortstack{First worked\\f.-dim. system}}}  &   \textbf{Qubit} &  \multicolumn{2}{c|}{\textbf{\shortstack{First conceptual \\ single-photon sys.}}}   \\ 
\cline{2-5} \cline{7-8}
{} & example & chapter  & example & chapter &   &  example & chapter  \\ 
\hline
\midrule

\textbf{Feynman '11 ('65)} & free p. & 16/21 & spin-1 & \cellcolor{blue!25}5/21 & / & pol. & 11/21 \\ 
Dirac '67$^4$ & free p. & 5/12 & spin-1/2 & 6/12 & / & pol. & \cellcolor{blue!25}1/12 \\ 
Shankar '94$^2$  & free p. & \cellcolor{blue!25}5/21 & spin-1/2 & 14/21 & / & / & /  \\ 
Messiah '95$^2$ & inf. sq. well & \cellcolor{blue!25}3/21 & spin-1/2 & 13/21 & / & / & / \\ \hline
Auletta '09$^1$ & inf. sq. well & 3/17 & MZI & \cellcolor{blue!25}1/17 &  17/17 & MZI & \cellcolor{blue!25}1/17    \\ 
Cohen-T. '20$^2$ & free p. & \cellcolor{blue!25}1/21 & pol. & 3/21 & /  & pol. &  3/21 \\ 
Zwiebach '22$^1$ & free p. & 4/30 & MZI & 2/30 & 18/30  & pol. & \cellcolor{blue!25}1/30 \\ 
Cetto '26$^1$ & sq. well & \cellcolor{blue!25}5/19 & spin-1/2 & 14/19 & 14/19 & / & / \\ 
Freericks '26$^1$ & free p. & 7/23 & spin-1/2 & 6/23 & 10, 14/23 & MZI & \cellcolor{blue!25}4/23    \\ 
\midrule

Liboff '03$^4$ & free p. & \cellcolor{blue!25}3/16 & spin-1/2 & 11/16 & 16/16 & Bell test & 11/16 \\ 
Binney '14$^1$ & free p. & 2/13 & ammonia & 5/13 & 6/13 & pol. & \cellcolor{blue!25}1/13  \\ 
Dunningham '18$^2$ & inf. sq. well & 4/13 & two-state & 4/13 & 12/13 & MZI &\cellcolor{blue!25}1/13  \\ 
Griffiths '18$^3$ & inf. sq. well & \cellcolor{blue!25}2/12 & spin-1/2 & 4/12 & / & / & /  \\ 
\textbf{Lvovsky '18$^1$} & free p. & 3/5 & pol. & \cellcolor{blue!25}1/5 & \cellcolor{blue!25}1/5 & pol. & \cellcolor{blue!25}1/5 \\ 
\textbf{Pade '18$^2$} & inf. sq. well & 5/28 & pol. & \cellcolor{blue!25}2/28 & 26/28  & pol. & \cellcolor{blue!25}2/28  \\ 
Banks '19$^1$ & free p. & 3/21 & ammonia & \cellcolor{blue!25}2/21 & 20$^\dagger$/21 & / & / \\ 
McIntyre '22$^1$ & inf. sq. well & 5/16 & spin-1/2 &\cellcolor{blue!25}1/16 &  16/16 & / & /  \\ 
\textbf{Basdevant '23$^3$} & free p. & \cellcolor{blue!25}3/18 & ammonia & 7/18 & 18$^\dagger$/18 & pol.  & 8/18  \\ 
\textbf{Kok '23$^2$} & inf. sq. well & 8/11  & spin-1/2 &  3/11 & 6/11 & MZI & \cellcolor{blue!25}1, 2/11  \\ 
Larkoski '23$^1$ & inf. sq. well & \cellcolor{blue!25}5/13 & spin-1/2 & 8/13 & 12/13 & HOM & 12/13   \\ 
\textbf{Styer '24} & inf. sq. well & 7/24 & spin-1/2 & \cellcolor{blue!25}1/24 & \cellcolor{blue!25}1/24 & pol. &  2/24 \\ 
\midrule

\cellcolor{gray!15} Haroche '06$^1$ & h. o. & 3/10 & two-state & \cellcolor{blue!25}2/10 &  \cellcolor{blue!25}2/10 & pol. & \cellcolor{blue!25}2/10 \\ 
\cellcolor{gray!15} Beck '12$^1$ & barrier & 11/17 & pol. & \cellcolor{blue!25}3, 4, 5/17 & 17/17 & pol. & \cellcolor{blue!25}3/17 \\ 
\cellcolor{gray!15} Townsend '12$^2$ & free p. & 6/14 & spin-1/2 & \cellcolor{blue!25}1, 2/14  & / & pol. & 2/14  \\ 
\cellcolor{gray!15} Dick '20$^3$ & barrier & \cellcolor{blue!25}3/23 & spin-1/2 & 8/23 & / & / & /  \\ 
\cellcolor{gray!15} Simon '25$^1$ & atom & 4/27 & two-state& 3/27 & \cellcolor{blue!25}1/27 & pol. & \cellcolor{blue!25}1/27 \\ 
\midrule

\cellcolor{gray!45} Landau '77$^3$ & sq. well &  \cellcolor{blue!25}3/18 & spin-1/2  & 8/18 & / & / & / \\ 
\cellcolor{gray!45} \textbf{Jackson '24 ('89)}  & free p. & \cellcolor{blue!25}1/10 &  spin-1/2 & 2/10 & / & / & / \\ 
\cellcolor{gray!45} Gottfried '03$^2$  & interf.  &  \cellcolor{blue!25}1/13 & spin-1/2 &  3/13 &  4/13 & (HOM) & 1/13 \\ 
\cellcolor{gray!45} Commins '14$^1$ & free p. & 6/25 & pol. & \cellcolor{blue!25}3/25 & / & pol. & \cellcolor{blue!25}3/25  \\ 
\cellcolor{gray!45} Balletine '15$^2$ &  free p. & \cellcolor{blue!25}4/21 & spin-1/2 & 7/21 & 21/21 &  / & /  \\ 
\cellcolor{gray!45} Wan '19$^1$ & h. o.  & 35/37 & spin-1/2 & \cellcolor{blue!25}14/37 & / & / & / \\ 
\cellcolor{gray!45} Sakurai '21$^3$ & free p.  & 1.7/8  &  spin-1/2 &  1.3, 1.4/8 & / & pol. & \cellcolor{blue!25}1.1/8   \\ 
\cellcolor{gray!45} Năstase '23$^1$ & free p. & 5, 6/58 & spin-1/2 & \cellcolor{blue!25}4/58  & \cellcolor{blue!25}4, 33/58 & / & / \\ 
\cellcolor{gray!45} \textbf{Levy '26} & lattice field & 3/12 & two-state & \cellcolor{blue!25}1/12 &  \cellcolor{blue!25}1/12 & Bell test  & 10/12  \\ 
\cellcolor{gray!45} Martin '26$^1$ & free p. & 6/29 & spin-1/2 & \cellcolor{blue!25}1/29 & 26/29 & Bell test & 14/29 \\ 
\bottomrule

\end{tabular}}
 \begin{tablenotes}
\item \textit{Note:} $\dagger$ indicates use of the term \textit{Q-bit}.
\item \textit{Abbreviations:} free p.\ = free particle; inf.\ sq.\ well = infinite square well; pol.\ = photon polarization; HOM = Hong--Ou--Mandel interferometer; MZI = Mach--Zehnder interferometer.
 \end{tablenotes}
\end{threeparttable}
\end{table*}

\subsubsection{Physical entry points and their relation to mathematical sequencing}
Across our corpus, the choice and timing of initial physical systems broadly reflect the conventional spin-first/wave-first distinction, yet cross-referencing physical examples with the mathematical sequencing in Table~\ref{table:1} reveals that a textbook's first physical illustration, first worked example, and first mathematical formalism are distinct curricular choices that do not always align. For instance, photon polarization frequently serves as an alternative finite-dimensional entry point—sometimes even in texts that develop wave mechanics as their initial mathematical formalism. Conversely, several texts develop abstract Hilbert-space and operator tools first, but the first physical examples are continuous-variable systems.

\paragraph*{Authors' rationales for physical entry points}
Authors who favor finite-dimensional starting points emphasize their conceptual accessibility and relatively modest mathematical demands. While acknowledging that infinite-dimensional systems connect more naturally to classical mechanics and display key phenomena (e.g., tunneling and atomic spectra), these authors argue that continuous systems can obscure the revelation of basic quantum principles as well as the logical sequence between fundamental postulates and derived results.

\subsubsection{Choice and contextualization of illustrative systems}
Beyond the broad division between finite- and infinite-dimensional entry points, textbooks differ markedly in the specific physical realizations chosen to contextualize dynamics, measurement, and modern applications.

\paragraph{Infinite-dimensional systems.}
Among introductory texts, a recurring choice in selecting the first worked infinite-dimensional model is between the free particle (foregrounding continuous spectra and wave-packet propagation) and bound-state potentials such as the infinite or finite square well (foregrounding confinement, boundary conditions, and discrete energy levels, sometimes linked to nanoelectronics). Some texts bridge both routes by using the free particle to motivate wave-mechanical tools before solving the square well as the first boundary-value problem. Advanced texts exhibit greater diversity—opening with potential barriers, the harmonic oscillator (often deferred until after formal postulational development), or specialized systems aligned with the text's thematic focus (e.g., atomic systems, field oscillators, lattice fields, or two-particle interferometry). Even when selecting the same system, texts assign it different mathematical roles, such as treating the free particle and harmonic oscillator primarily through symmetry and algebraic methods rather than differential equations.

\paragraph{Finite-dimensional systems.}
While spin-$1/2$ and photon polarization dominate finite-dimensional introductions, several texts select alternative realizations that highlight different physical or pedagogical priorities: spin-$1$ (chosen to provide a three-state prototype for general quantum behavior), the ammonia molecule (connecting two-state dynamics to molecular physics and tunneling), or an abstract, realization-independent two-state model used to establish mathematical and conceptual foundations concurrently.

\paragraph{Qubits.}
The term \textit{qubit} appears explicitly in the vast majority of post-2000 textbooks in our corpus, but its contextualization follows two distinct patterns. Some texts—including later editions of established works—present qubits in self-contained, late chapters on quantum information, computation, and entanglement, even if generic two-state systems were introduced much earlier. By contrast, several recent texts—both QIST-focused and broadly scoped—adopt the qubit as an early, foundational organizing concept for quantum theory itself.

\paragraph{Single-photon systems.}
Although using photon polarization to introduce quantum states and measurement has a classic precedent in Dirac '67$^4$, many subsequent textbooks anchor their foundations almost exclusively in massive particles, omitting single-photon systems or restricting them to late chapters on Bell's theorem and foundations. Conversely, a subset of modern textbooks champions single-photon and quantum-optical systems—including polarization, Mach--Zehnder interferometers, Hong--Ou--Mandel interference, and photon bunching—as primary pedagogical vehicles. Authors' rationales highlight three advantages: continuity with classical polarization optics, mathematical accessibility, and direct connection to realistic tabletop laboratory experiments and modern quantum technologies—arguing that these conceptual and experimental benefits outweigh the fact that a full theoretical treatment of the photon requires relativistic quantum field theory.

\subsection{Theoretical Concepts: Postulates, Principles, Foundation and Interpretations}
\label{sec:res_3}
PER literature indicates that how instruction addresses foundational and interpretive issues shapes students' conceptual understanding of QM and their views on the nature of quantum theory \cite{Baily_2015, Dubson_2009}. One key source of curricular variation is whether formal postulates are taught explicitly and how early they are introduced \cite{Dubson_2009, Kruse_2025}, yet relatively few studies concern the postulates themselves.

In the broader textbook literature, authors adopt contrasting positions on the pedagogical role of formal axioms—reflecting competing pedagogical trade-offs between deductive coherence and inductive motivation (informing RQ2.1, RQ2.2). 
Some authors advocate building QM systematically from its postulates analogous to Newton's laws, though scaffolded historically due to their counterintuitive nature \cite{Puri_2017}, whereas others view postulates as summaries of experimental facts and favor physically motivating the need for quantum theory before introducing its axiomatic structure \cite{Bohm_1986, Bohm_1989}. 
These differing philosophies lead to distinct choices regarding the selection, sequencing, and contextualization of foundational rules.

Although there is no universal consensus on the exact number, order, or wording of QM's foundational statements, there is broad agreement on the theory’s core rules. In our topic-level analysis (RQ1), we examine 8 foundational postulates---\textbf{state, observable, measurement outcome, measurement probability, post-measurement state, time evolution, composite systems, and symmetrization}---and 6 physical principles: 5 traditional principles (\textbf{superposition, complementarity, uncertainty, correspondence, and Pauli's exclusion principle}) together with a modern information-theoretic principle (\textbf{the no-cloning theorem}). This selection is not intended as an exhaustive or prescriptive account of quantum foundations: it excludes principles not specific to QM (such as the action principle) as well as proposals with limited uptake in standard textbooks (such as Zeilinger’s principle of information quantization \cite{Zeilinger_1999}). Moreover, the foundational status of several included concepts---such as wave--particle duality, complementarity, and correspondence---remains actively contested. Our categories therefore serve as an analytical baseline for comparing recurring foundational statements and the debates surrounding their pedagogical status.

A closely related source of divergence is the treatment of quantum interpretation and, in particular, the measurement process, where differing instructional emphases produce measurable differences in student thinking \cite{Baily_2015}. In the broader textbook literature, some authors prefer postponing interpretive questions until students have developed practical command of the formalism, whereas others argue that measurement and interpretation are essential to understanding the theory from the outset \cite{schwindt_2016, peebles_2020, Trachanas_2017}. Simultaneously, experimental advances and the rise of QIST have elevated the prominence of topics that bridge mathematical structure, laboratory practice, and philosophical interpretation while underpinning modern quantum technologies. To capture how textbooks navigate these developments and define their broader pedagogical orientations (RQ1, RQ2.3), we track 5 topics in foundations and interpretations: \textbf{entanglement, nonlocality, interpretations of QM, the measurement problem, and decoherence}.

Across our 36-textbook corpus, many texts establish a conceptual basis for QM through dedicated opening chapters---often signaled by dedicated opening chapters with titles such as ``Fundamental Concepts'' or ``Conceptual Ideas''---yet content analysis reveals substantial variation in their underlying pedagogical choices across foundational postulates (Sec.~\ref{sec: res_3.1}), physical principles (Sec.~\ref{sec: res_3.2}), and selected topics in foundations and interpretations (Sec.~\ref{sec: res_3.3}).

\subsubsection{Foundational postulates of QM}
\label{sec: res_3.1}
Building on earlier comparative compilations of QM postulates (such as the appendix in Pade '18$^2$), Table~\ref{table:3.1} summarizes the selection and sequencing of 8 foundational rules across our corpus.

\begin{table*}[h!]
\caption{Chapter-level presentation of fundamental postulates and rules of QM in the selected texts. Shading identifies chapters in which at least five of the recorded foundational statements are introduced together: blue denotes statements explicitly labeled as a \textit{postulate}, \textit{rule}, \textit{principle}, or \textit{theorem}, while green denotes corresponding statements presented without these labels. Red shading identifies an exceptional organization in which individual postulates receive dedicated chapters.}
\label{table:3.1}
\begin{threeparttable}
\centering
\footnotesize
\renewcommand{\arraystretch}{1.2}
\resizebox{1.0\textwidth}{!}{
\begin{tabular}{|l|cccccc|cc|}
\toprule
Text ID & \multicolumn{6}{c|}{Postulates / Rules of QM} & \multicolumn{2}{c|}{for many-body QM}\\ \midrule
{} & \textbf{State} & \textbf{Observable} & \multicolumn{3}{c}{\textbf{Measurement}} & \textbf{\shortstack{Time\\ evolution}} & \textbf{\shortstack{Composite \\ system}} & \textbf{Symmetrization} \\ 
&&& \textbf{outcome} & \textbf{probability} & \textbf{state after} & & & \\ \hline
\midrule

\textbf{Feynman '11 ('65)} & (20)/21 & (20)/21 & (5), (20)/21 & (3)/21 & (5)/21 & (8), (16)/21 & (\{12\})/21  & (4)/21  \\ 
Dirac '67$^4$ & \cellcolor{blue!25}2/12 & \cellcolor{blue!25}2/12 & \cellcolor{blue!25}2/12 & \cellcolor{blue!25}2/12 & \cellcolor{blue!25}2/12 & \{5\}/12 & (\{9\})/12  & [(9)]/13  \\ 
Shankar '94$^2$  & \cellcolor{blue!25}4/21 & \cellcolor{blue!25}4/21 & \cellcolor{blue!25}4/21 & \cellcolor{blue!25}4/21 & \cellcolor{blue!25}4/21 & \cellcolor{blue!25}4/21 & ([10])/21 & [10]/21   \\ 
Messiah '95$^2$ & 2, 7/21 & 5, 7/21 & 5, 8/21 & 5, 8/21 & 8/21 & \{8\}/21 & (7)/21 & 14/21   \\ 
Auletta '09$^1$ & (1)/17  & 2$^\dagger$/17 &  2$^\dagger$/17 & ([1]), 2$^\dagger$/17 & (1)/17 & \{(3)\}/17 &  (6)/17  &  [7]/17   \\ 
Cohen-T. '20$^2$ & \cellcolor{blue!25}3/21 & \cellcolor{blue!25}3/21 & \cellcolor{blue!25}3/21 & \cellcolor{blue!25}3/21 & \cellcolor{blue!25}3/21 &  \cellcolor{blue!25}3/21 & (3)/21 & 14/21   \\ 
Zwiebach '22$^1$ & \cellcolor{blue!25}16/30 & \cellcolor{blue!25}16/30 & \cellcolor{blue!25}16/30 & \cellcolor{blue!25}[3], 16/30 & \cellcolor{blue!25}16/30 & \cellcolor{blue!25}\{16\}/30 & \cellcolor{blue!25}16/30 & \cellcolor{blue!25}16/30   \\ 
Cetto '26$^1$ & \cellcolor{blue!25}(3), 4/19 & \cellcolor{green!25}(3), (4)/19 & \cellcolor{green!25}(4)/19 & \cellcolor{green!25}[4]/19 & \cellcolor{green!25}(4)/19 & 7/19 & (14)/19 & [15]/19   \\ 
Freericks '26$^1$ & (5)/23 & (5)/23 & 6, 7/23 & [3], 7, [13]/23 & 7/23  & \{14\}/23 & (6)/23  & (17)/23     \\ 
\midrule

Liboff '03$^4$ & \cellcolor{blue!25}3/16 & \cellcolor{blue!25}3/16 & \cellcolor{blue!25}3/16 & [2]/16 & \cellcolor{blue!25}3/16  & \cellcolor{blue!25}3/16 & (\{3\})/16 & 12/16  \\ 
Binney '14$^1$ &  1/13 & (2)/13 & (1)/13 & [1]/13 & (1)/13  & (2)/13 & (\{6\})/13 & 11$^\dagger$/17  \\ 
Dunningham '18$^2$ & \cellcolor{blue!25}3, 8/16 & \cellcolor{blue!25}3, 8/16 & \cellcolor{blue!25}8/16 & \cellcolor{blue!25}[3], 8/16 & \cellcolor{blue!25}8/16 & \cellcolor{blue!25}3, 8/16 & (12)/16 & [6]/16   \\ 
Griffiths '18$^3$ & \cellcolor{green!25}(3)/12 & \cellcolor{green!25}(3)/12 & \cellcolor{green!25}(3)/12 & \cellcolor{green!25}\{3\}/12 & \cellcolor{green!25}(3)/12 &  (1)/12 &  (\{4\})/12 & 5$^\dagger$/12   \\ 
\textbf{Lvovsky '18$^1$} & \cellcolor{blue!25}1/5 & \cellcolor{green!25}(1)/5 & \cellcolor{green!25}(1)/5 & \cellcolor{blue!25}[1], [2], [5]/5 & \cellcolor{blue!25}1/5 & \cellcolor{blue!25}\{1\}/5 & (2)/5 & [(4)]/4   \\ 
\textbf{Pade '18$^2$} &  \cellcolor{blue!25}14/28 & \cellcolor{blue!25}14/28 & \cellcolor{blue!25}14/28 & \cellcolor{blue!25}[14]/28 & \cellcolor{blue!25}14/28 & \cellcolor{blue!25}14/28 & (20)/28 & [23]/28   \\ 
Banks '19$^1$ & (1)/21 & (2)/21 & (1)/21 & (1), [2]/21 & (2)/21 & \{(2)\}/21 & (6)/21 &  (15)/21  \\ 
McIntyre '22$^1$ & \cellcolor{blue!25}1/16 & \cellcolor{blue!25}1/16 & \cellcolor{blue!25}1/16 & \cellcolor{blue!25}1/16 & \cellcolor{blue!25}1/16 & \cellcolor{blue!25}1/16 & (\{4\})/16 &  13/16  \\ 
\textbf{Basdevant '23$^3$} & \cellcolor{blue!25}6$^\dagger$/19 & \cellcolor{blue!25}6$^\dagger$/19 & \cellcolor{blue!25}6$^\dagger$/19 & \cellcolor{blue!25}6$^\dagger$/19 & \cellcolor{blue!25}6$^\dagger$/19 & \cellcolor{blue!25}6$^\dagger$/19 & (9)/19 & (15$^\dagger$)/19   \\ 
\textbf{Kok '23$^2$} &  \cellcolor{blue!25}5/11 & \cellcolor{blue!25}5/11 & \cellcolor{blue!25}5/11 & \cellcolor{blue!25}[2], [5]/11 & \cellcolor{blue!25}5/11 & \cellcolor{blue!25}5/11 & (\{6\})/11 & [Epilogue]   \\ 
Larkoski '23$^1$ & 4/13 & 4/13 & 4/13 & [3]/13 & 13/13 & \{(4)\}/13 & (12)/13 & [(6)]/13   \\ 
\textbf{Styer '24} & \cellcolor{blue!25}4/24 & \cellcolor{blue!25}4/24 & \cellcolor{blue!25}4/24 & \cellcolor{blue!25}4/24 & \cellcolor{blue!25}4/24 & \{5\}/24 & (\{6\})/24  & [15]/24  \\ 
\midrule

\cellcolor{gray!15} Haroche '06$^1$ & \cellcolor{green!25}(2)/20 & \cellcolor{green!25}(2)/10 & \cellcolor{green!25}(2)/10 & \cellcolor{blue!25}2/10 & \cellcolor{blue!25}2/10 &  \cellcolor{green!25}(2)/10 & \cellcolor{green!25}(2)/10 &  \cellcolor{blue!25}2/10  \\ 
\cellcolor{gray!15} Beck '12$^1$ & \cellcolor{blue!25}(3), 5/17 & \cellcolor{blue!25}(4), 5/17 & \cellcolor{blue!25}5/17 & \cellcolor{blue!25}[5], [8]/17 & \cellcolor{blue!25}5/17 & \{(9)\}/17 &  ([8])/17 &  [13]/15  \\ 
\cellcolor{gray!15} Townsend '12$^2$ & (1)/14  & (2)/14 & (1)/14 & (1), \{6\}/14 & (2)/14 & \{(4)\}/14 & ([5])/14 & [12]/14   \\ 
\cellcolor{gray!15} Dick '20$^3$ & (4)/21 & (4)/21 & (5)/21 & \{1\},(5)/21  & (13)/21 & (13)/21 & (8)/21 &  (17)/21   \\ 
\cellcolor{gray!15} Simon '25$^1$ & \cellcolor{green!25}(3)/27 & \cellcolor{green!25}(3)/27 & \cellcolor{green!25}(3)/27 & \cellcolor{green!25}[3]/27 & \cellcolor{green!25}(3)/27 & (assumed) & \cellcolor{green!25}(3)/27 & [(6)]/27  \\ 
\midrule

\cellcolor{gray!45} Landau '77$^3$ & \cellcolor{green!25}(1)/18 & \cellcolor{green!25}(1)/18 & \cellcolor{green!25}(1)/18 & \cellcolor{green!25}(1)/18 & \cellcolor{green!25}(1)/18 & \{3\}/18 & (\{9\})/18 & 9$^\dagger$/18 \\ 
\cellcolor{gray!45} \textbf{Jackson '24 ('89)}  & \cellcolor{blue!25}2/10 & \cellcolor{blue!25}2/10 & \cellcolor{blue!25}2/10 & \cellcolor{blue!25}2/10 & \cellcolor{blue!25}2/10 & \cellcolor{blue!25}\{2\}/10 & ([4])/10 & [6]/10   \\ 
\cellcolor{gray!45} Gottfried '03$^2$  & \cellcolor{blue!25}2/13 & \cellcolor{blue!25}2/13 & \cellcolor{blue!25}2/13 & \cellcolor{blue!25}2/13 & 12/13 & \cellcolor{blue!25}\{2\}/13 & ([1])/13 & [(6)]/13  \\ 
\cellcolor{gray!45} Commins '14$^1$ & \cellcolor{blue!25}3/25 & \cellcolor{blue!25}3/25 & \cellcolor{blue!25}3/25 & \cellcolor{blue!25}3/25 &  \cellcolor{blue!25}3/25 & \cellcolor{blue!25}3/25 & (assumed) &  11/25 \\ 
\cellcolor{gray!45} Balletine '15$^2$ &  \cellcolor{blue!25}2$^*$/21 & \cellcolor{blue!25}2/21 & \cellcolor{green!25} (2)/21 & \cellcolor{green!25}(2)/21 & \cellcolor{green!25}(2)/21 & \{4\}/21 & (3°)/21 & 17/21   \\ 
\cellcolor{gray!45} Wan '19$^1$ & \cellcolor{red!25}25/37 & \cellcolor{red!25}26, 27/37 & \cellcolor{red!25} 26/37 & \cellcolor{red!25} 28/37  & \cellcolor{red!25} 30/37 & \cellcolor{red!25} 29/37 & (33/37) & [(33)]/37  \\ 
\cellcolor{gray!45} Sakurai '21$^3$ & \cellcolor{blue!25}1/8 & \cellcolor{green!25}(1)/8 & \cellcolor{green!25}(1)/8 & \cellcolor{blue!25}1/8 & \cellcolor{green!25}(1)/8 & \{(2)\}/8 & (3)/8 & 7/8   \\ 
\cellcolor{gray!45} Năstase '23$^1$ & \cellcolor{blue!25}3/58 & \cellcolor{blue!25}3/58 & \cellcolor{blue!25}3/58 & \cellcolor{blue!25}3/58 & \cellcolor{blue!25}3/58 & \cellcolor{blue!25}3/58 & (3)/58 & [20]/58  \\ 
\cellcolor{gray!45} \textbf{Levy '26} & \cellcolor{blue!25}1/12 & \cellcolor{blue!25}1/12 & \cellcolor{blue!25}1/12 & \cellcolor{blue!25}[1]/12  & \cellcolor{blue!25}1/12  &  \cellcolor{blue!25}\{1\}/12 & (2)/12 & ([3])/12   \\ 
\cellcolor{gray!45} Martin '26$^1$ & \cellcolor{blue!25}3/29 & \cellcolor{blue!25}3/29 & \cellcolor{blue!25}3/29 & \cellcolor{blue!25}[3]/29 & \cellcolor{blue!25}3/29 & \cellcolor{blue!25}3/29 & (2)/29 & [18]/29    \\ 
\bottomrule

\end{tabular}
}
 \begin{tablenotes}
\item \textit{Note:} Parentheses \textbf{(~)} indicate discussions where the explicit designation (\textit{postulate}, \textit{rule}, \textit{principle}, or \textit{theorem}) is omitted.
\item Special markers: $\dagger$ indicates formulation as a \textit{principle}; $\star$ indicates a postulate formulated as a postulate on the state operator.
\item Brackets and symbols carry column-specific meanings: 
\textbf{Measurement outcomes:} \textbf{[~]} = explicit mention of \textit{Born's rule}, \textbf{\{~\}} = \textit{Born's  interpretation}; 
\textbf{Time evolution:} \textbf{\{~\}} = Schrödinger equation derived from a fundamental postulate on unitary evolution; 
\textbf{Composite systems:} \textbf{[~]} = defined via the direct product, \textbf{\{~\}} = defined via product, $^\circ$ = defined via the Kronecker product; 
\textbf{Symmetrization:} \textbf{[~]} = presentation of the spin--statistics theorem without explicit mention of the symmetrization postulate.
\item \textit{Assumed} indicates that a topic is utilized without a substantive introduction and is treated as prior knowledge.
 \end{tablenotes}
\end{threeparttable}

\end{table*}

\paragraph{Formulation, placement, and authors' rationales.}
Introductory textbooks vary widely in how they organize the postulates: some state them explicitly at the outset, others consolidate them only after substantial phenomenological or mathematical preparation, and still others develop the underlying rules without collecting them into a formally labeled postulate set. By contrast, graduate-level texts more consistently present an explicit axiomatic formulation near the beginning of the book, whereas advanced-undergraduate bridge texts frequently present foundational content without explicit postulate labels. Terminology and structural granularity also vary: some texts label these rules ``principles'' or ``statistical algorithms'' rather than ``postulates,'' some distinguish an initial wave-mechanical formulation from a later general Hilbert-space formulation, and others distribute individual postulates across multiple dedicated chapters.

Authors justify these structural choices by weighing axiomatic coherence against pedagogical accessibility. Authors who introduce postulates early often anchor them immediately in concrete two-state systems (such as photon polarization or Stern--Gerlach setups) or precede them with a dedicated mathematical chapter. Conversely, authors who defer or avoid an explicitly axiomatic opening argue that unmotivated postulates can appear arbitrary and cumbersome to beginners, preferring either to build rules inductively from experimental evidence or to use a later postulational chapter to synthesize prior analytical and algebraic developments.

\paragraph{Variations among individual postulates.}
Across the 8 tracked postulates, notable variations in contextualization and scope emerge:
\begin{itemize}
    \item \textit{States and observables:} While most texts formulate state postulates in terms of pure state vectors, a few define the state operator (density matrix) as the primary foundational concept from the outset or add an explicit postulate for canonical quantization mapping classical observables to quantum operators.
    \item \textit{Measurement outcomes, probabilities, and post-measurement states:} Texts differ in whether measurement axioms foreground eigenvalues or expectation values, whether the Born rule is named explicitly or framed as a ``generalized statistical interpretation,'' and how state collapse (or the projection postulate) is handled. Because of its interpretational sensitivity, some texts separate the post-measurement state rule from the other postulates, deferring it to later chapters on measurement theory or treating it as implicit in reduced density-matrix dynamics.
    \item \textit{Time evolution, composite systems, and symmetrization:} Some texts present the time-dependent Schrödinger equation itself as a fundamental postulate, others postulate unitary time evolution first and derive the Schrödinger equation as a consequence, and some omit an explicit time-evolution postulate altogether. Similarly, while tensor-product structures for composite systems are typically introduced implicitly as mathematical tools, Zwiebach '22$^1$ elevates composite systems to an independent postulate. Finally, some texts explicitly present symmetrization as an independent postulate, whereas others discuss the spin–statistics theorem. Some authors regard the spin–statistics connection as a formal postulate necessary within non-relativistic QM, whereas others critique postulating Bose and Fermi statistics and instead emphasize their derivation from quantum field theory.
\end{itemize}

\subsubsection{Foundational principles of QM}
\label{sec: res_3.2}
Table~\ref{table:3.2} summarizes the selection and sequencing of 6 foundational statements commonly referred to as principles.

\begin{table*}[h!]
\caption{Chapter-level presentation of core foundational principles
of QM. Shading identifies chapters in which at least three of the
recorded principles are introduced: blue denotes statements explicitly
labeled as \textit{principle}, \textit{rule}, or \textit{theorem}, while green denotes corresponding content presented without these labels.} 
\label{table:3.2}
\begin{threeparttable}
\centering
\footnotesize
\renewcommand{\arraystretch}{1.2}
\resizebox{1.0\textwidth}{!}{
\begin{tabular}{|l|ccccc|c|}
\toprule
{Text ID} &  \multicolumn{5}{c|}{Principles} &  Theorem \\ \midrule
{} & \textbf{Superposition} & \textbf{Complementarity} & \textbf{Uncertainty} & \textbf{Correspondence} & \textbf{Exclusion} & \textbf{No-Cloning}  \\ \hline
\midrule

\textbf{Feynman '11 ('65)} &  (3)/21 & / & 1/21 & / & 4/21 & /  \\ 
Dirac '67$^4$ & 1/12 & / & 4/12 & / & 9/12 & /   \\ 
Shankar '94$^2$  & 4/21 & [3]/21 & 4/21 & 7/21 & 10/21 & /   \\ 
Messiah '95$^2$ & 7/21 & 4/21 & [4]/21 & 1/21 & 14/21 & /   \\ 
Auletta '09$^1$ & 1/17 & 1/17 & 2/17 & 2/17 & 7/17 & 15/17    \\ 
Cohen-T. '20$^2$ & 3/21 & (1)/21 & 1, 3/21 &  / & 14/21 & /    \\ 
Zwiebach '22$^1$ & (1)/30 & / & 1, 4, 15/30 & 7/30 & 21/30 & 18/30   \\ 
Cetto '26$^1$ & 4/19 & / & [3], [8]/19 & 4/19 & 15/19 & /   \\ 
Freericks '26$^1$ & 2/23 & 3/23 & 7/23 & 11/23 & 10/23 & 6/23     \\ \hline
\midrule

Liboff '03$^4$ & 5/16 & (2)/16 & 2/16 & 4/16 & 12/16 & /  \\ 
Binney '14$^1$ & (1)/13 & / & 2/13 & 2/13 & 11/13  &   / \\ 
Dunningham '18$^2$ & \cellcolor{blue!25}3/13 & \cellcolor{blue!25}3/13 & \cellcolor{blue!25}3/13 & 8/13 & 6/13 & (12)/13   \\ 
Griffiths '18$^3$ & (6)/12 & 12/12 & 1, 3/12 & 11/12 & 5/12 & 12/12   \\ 
\textbf{Lvovsky '18$^1$} & \cellcolor{green!25} (1)/5 & \cellcolor{blue!25}1/5 & \cellcolor{blue!25}1/5 & \cellcolor{blue!25}1/5 & 4/5 & 2/5 (ex.)   \\ 
\textbf{Pade '18$^2$} & 1/28 & (6)/28 & 6/28, App. I & 3/28 & 23/28 & 26/28    \\ 
Banks '19$^1$ & (2)/21 & [5]/21 & 4, 6/21 & / & 5/21 &  20/21  \\ 
McIntyre '22$^1$ & (1)/16 & (6)/16 & 2, 6, 9/16 & / & 13/16 & /   \\ 
\textbf{Basdevant '23$^3$} & \cellcolor{blue!25}3, 6/19 & \cellcolor{green!25} [3]/19 & \cellcolor{blue!25}3, 9, 16/19 & 4/19 & 15/19 & 18/19    \\ 
\textbf{Kok '23$^2$} &  (2)/11 & 11/11 & 9/11 & 11/11 & 8/11 & 6/11   \\ 
Larkoski '23$^1$ & / & / & 4/13 & 4/13 & / & /   \\ 
\textbf{Styer '24} & (2)/24 & [2]/24 & [4], [6]/24 & 15/24 &  /  & /  \\ 
\midrule

\cellcolor{gray!15} Haroche '06$^1$ & \cellcolor{blue!25}2/10 & \cellcolor{blue!25}2/10 & \cellcolor{blue!25}2/10 & 3/10 & \cellcolor{blue!25}2/10  & \cellcolor{blue!25}2/10    \\ 
\cellcolor{gray!15} Beck '12$^1$ & (3)/17 & 5/17 & [5], [10]/17 & / & 13/17  & 17/17    \\ 
\cellcolor{gray!15} Townsend '12$^2$ & (1)/14  & / & [3], [4], 6/14 & 7/14 & 12/14 & 5/14   \\ 
\cellcolor{gray!15} Dick '20$^3$ & / & / & 5/21 & 18/21 & (assumed) & /   \\ 
\cellcolor{gray!15} Simon '25$^1$ & (1), (3)/27 & / & 3/27 & / & 5/27 & 12/27  \\ 
\midrule

\cellcolor{gray!45} Landau '77$^3$ & 1/18 & / & 1/18 & (1)/18 & 9/18 & /   \\ 
\cellcolor{gray!45} \textbf{Jackson '24 ('89)}  & (2)$^\dagger$/10 & / & 2/10 & 3/10 & 6/10 & /   \\ 
\cellcolor{gray!45} Gottfried '03$^2$  & \cellcolor{blue!25}1/13 & \cellcolor{blue!25}1/13 & \cellcolor{blue!25}1/13 & \cellcolor{blue!25}1/13 & 6/13 &  /  \\ 
\cellcolor{gray!45} Commins '14$^1$ & 3/25  & [1]/25  & 1, 4/25 & / & 11/25 &  / \\ 
\cellcolor{gray!45} Balletine '15$^2$ & (assumed)  & (0)/21 & [8]/21 & / & 17/21 & 8,  21/21   \\ 
\cellcolor{gray!45} Wan '19$^1$ & 5/37 & [34]/37 & [22], [28]/37 & / & 33/37 & /  \\ 
\cellcolor{gray!45} Sakurai '21$^3$ & 2/8 & / & 1/8 & (2)/8 & 7/8 & /   \\ 
\cellcolor{gray!45} Năstase '23$^1$ & (assumed) & (0)/58 & 6/58 & 0/58 & 20/58 & 33/58    \\ 
\cellcolor{gray!45} \textbf{Levy '26} & 1/12 & / & (assumed) & (12)/12 & (assumed) & 10/12    \\ 
\cellcolor{gray!45} Martin '26$^1$ & 2/29 & [1]/29 & 3/29 & 4/29 & 12, 18/29 & 27/29    \\ 
\bottomrule

\end{tabular}}

 \begin{tablenotes}
\item \textit{Note:} Parentheses \textbf{(~)} indicate discussions where the exact term is absent, or where the explicit designation (\textit{rule}, \textit{principle}, or \textit{theorem}) is omitted.
\item A dagger ($\dagger$) indicates that the concept is formulated as an \textit{assumption}.
\item Square brackets \textbf{[~]} carry column-specific meanings: under \textbf{Complementarity}, \textbf{[~]} indicates coverage limited to wave--particle duality; under \textbf{Uncertainty}, \textbf{[~]} indicates alternative terminology (e.g., \textit{Heisenberg indeterminacy relation}, \textit{uncertainty relation}, or \textit{Heisenberg inequality}).
\item \textit{Abbreviations:} ex.\ = exercise; App.\ = appendix.
\item \textit{Assumed} indicates that a topic is utilized without a substantive introduction and is treated as prior knowledge.
 \end{tablenotes}
\end{threeparttable}
\end{table*}

\paragraph{Formulation and placement.}
Unlike postulates, these physical principles are typically distributed across multiple chapters as specific mathematical tools or physical systems arise. Only a few texts concentrate nearly all tracked principles into a single opening chapter—either integrating principles and postulates concurrently to launch experimental quantum physics, or using a qualitative overview of physical principles on physical principles to scaffold a subsequent chapter on formal axioms. Other texts embed principles within specific wave-mechanical or historical narratives.

\paragraph{Variations among individual principles.}
Individual principles receive markedly different degrees of foundational status and contextualization:
\begin{itemize}
    \item \textit{Superposition and uncertainty:} Superposition is often framed either mathematically (identifying state vectors in Hilbert space with linear superposition) or operationally (through spin or optical measurement outcomes, carefully distinguished from statistical mixtures). The uncertainty principle is the most universally included principle in the corpus, but ranges from specific position--momentum treatments and multi-stage pedagogical revisitations to generalized Robertson--Schrödinger derivations and philosophical distinctions between epistemic ``uncertainty'' and intrinsic ``indeterminacy.''
    \item \textit{Complementarity and correspondence:} Both principles exhibit wide divergence in placement and perceived validity. While several texts foreground them in opening chapters, some relegate them to late historical notes or footnotes, explicitly questioning their generality or criticizing conventional presentations of wave--particle duality in favor of field-theoretic accounts. Conversely, a few texts systematically distinguish multiple historical and mathematical forms of correspondence (e.g., Bohr, Dirac, and Weyl rules).
    \item \textit{Exclusion and no-cloning:} Pauli's exclusion principle is predominantly contextualized within later chapters on identical particles and state symmetrization, though a few texts introduce multi-particle concepts much earlier or frame exclusion as a historical precursor to general symmetry requirements. Where included, the no-cloning theorem is usually tied to quantum-information applications (teleportation and cryptography)—even when sequenced early—though some texts frame it more fundamentally in terms of state preparation or the non-objective reality of the wavefunction.
\end{itemize}

\subsubsection{Selected topics in foundations and interpretations}
\label{sec: res_3.3}

Table~\ref{table:3.3} summarizes the selection and sequencing of 5 selected topics in foundations and interpretations across the corpus.

\begin{table*}[h!]
\caption{Chapter-level presentation of selected topics in the foundations and interpretation of QM. Blue shading marks chapters in which all five selected topics are introduced.}
\label{table:3.3}
\begin{threeparttable}
\centering
\footnotesize
\renewcommand{\arraystretch}{1.2}
\resizebox{1.0\textwidth}{!}{
\begin{tabular}{|l|cc|ccc|}
\toprule
Text ID &  \textbf{Entanglement} & \textbf{Nonlocality} & \textbf{\shortstack{Interpretations \\ of QM}} & \textbf{\shortstack{Measurement \\ problem}}  &  \textbf{Decoherence}  \\ \hline
\midrule

\textbf{Feynman '11 ('65)} & / & / & / & / & /  \\ 
Dirac '67$^4$ & / & / & / & / & /     \\ 
Shankar '94$^2$  & / &  / & / & / & /   \\ 
Messiah '95$^2$ & / & / & / & / & /   \\ 
Auletta '09$^1$ & 16, 17/17 & 16/17 & 9/17 & 9/17   & 9/17   \\ 
Cohen-T. '20$^2$ & 21/21 & 21/21 & / &  / & 21/21   \\ 
Zwiebach '22$^1$ & 1, 18/30 & (18)/30 & 22/30 & [22]/30 & 22/30      \\ 
Cetto '26$^1$ & 14/19 & 4, 14/19 & 1/19 & 1(mentioned)/19  & 15/19    \\ 
Freericks '26$^1$ & 2, 6, 9/23  & 2/23 & 7, 23/23 & 23/23 & 22, 23/23      \\ 
\midrule

Liboff '03$^4$ & 16/16 & 11/16 & / & / & /   \\ 
Binney '14$^1$ & 6/13 & / & 6$^*$/13 & (6)/13  & 6/13     \\ 
Dunningham '18$^2$ & 12/13 & 12/13 & 3/13 & [3]/13 & /  \\ 
Griffiths '18$^3$ & 5/12 & 12/12 & 1/12 & [12]/12 & 12/12   \\ 
\textbf{Lvovsky '18$^1$} & \cellcolor{blue!25}2/5 & \cellcolor{blue!25}2/5  & \cellcolor{blue!25}2/5 & \cellcolor{blue!25}[2]/5 & \cellcolor{blue!25}2, 5/5  \\ 
\textbf{Pade '18$^2$} & 20/28 & 20/28 & 28/28  & 24/28 & 24/28   \\ 
Banks '19$^1$ & 20/21 & 10/21 & App. A & [10]/21, (App. A) &  10/21, App. A    \\ 
McIntyre '22$^1$ & \cellcolor{blue!25}4/16 & \cellcolor{blue!25}4/16 & \cellcolor{blue!25}4$^*$/16  & \cellcolor{blue!25}[4]/16  & \cellcolor{blue!25}4/16     \\ 
\textbf{Basdevant '23$^3$} & 17/19 & 17/19 & 4$^*$/19 & [4]/19 & 4/19    \\ 
\textbf{Kok '23$^2$} &  6/11 & 11/11 & 11/11 & 9/11 & 7/11   \\ 
Larkoski '23$^1$ &  12/13 & (12)/13 & 13/13 & 13/13 & 13/13   \\ 
\textbf{Styer '24} & 1, 2/24 & / & / & / & /    \\ 
\midrule

\cellcolor{gray!15} Haroche '06$^1$ & \cellcolor{blue!25}2/10 & \cellcolor{blue!25}2/10 & \cellcolor{blue!25}2$^*$/10 & \cellcolor{blue!25}2/10 & \cellcolor{blue!25}2/10     \\ 
\cellcolor{gray!15} Beck '12$^1$ & 8/17 & 8/17 & / & [8] & 8, 17/17     \\ 
\cellcolor{gray!15} Townsend '12$^2$ & 5/14  &  5/14 & / & / & /   \\ 
\cellcolor{gray!15} Dick '20$^3$ & (17)/23 & / & / & / & /     \\ 
\cellcolor{gray!15} Simon '25$^1$ & 3, 13/27 & 10, 14, 24/27 & / & [3]/27 & 3/27   \\ 
\midrule

\cellcolor{gray!45} Landau '77$^3$ & / & /  & / & / & /    \\ 
\cellcolor{gray!45} \textbf{Jackson '24 ('89)}  & / & / & / & / &  /   \\ 
\cellcolor{gray!45} Gottfried '03$^2$  &  2/13 &  12/13 & 1$^*$, 12/13 & [12]/13 & /   \\ 
\cellcolor{gray!45} Commins '14$^1$ & 3/25 & 3/25 & 25/25 & 25/25 &  25/25 \\ 
\cellcolor{gray!45} Balletine '15$^2$ & 21/21 & 20/21 & / & 9/21 & 21/21   \\ 
\cellcolor{gray!45} Wan '19$^1$ & 33/37 & 34/37 & / & 34/37 & 34/37   \\ 
\cellcolor{gray!45} Sakurai '21$^3$ & (3)/8 & (3)/8 & / &  / & /   \\ 
\cellcolor{gray!45} Năstase '23$^1$ & 4, 40/58 & (30), 35/58  & 31/58 & [35]/38  & 35/58    \\ 
\cellcolor{gray!45} \textbf{Levy '26} & 2, 10/12 & 10/12 & 10/12 & / & 2/12    \\ 
\cellcolor{gray!45} Martin '26$^1$ & 2, 14, 25/29 & 14/29 & 3/29 & 3, [26]/29 & 26/29    \\ 
\bottomrule

\end{tabular}
}
 \begin{tablenotes}
\item \textit{Note:} Parentheses \textbf{(~)} indicate discussion without explicit use of the term.
\item \textbf{Interpretations of QM:} $\ast$ indicates discussion focused mainly on the Copenhagen interpretation; 
\item \textbf{Measurement problem:} \textbf{[~]} indicates discussion limited to Schrödinger's cat paradox.
 \end{tablenotes}
\end{threeparttable}
\end{table*}

\paragraph{Inclusion, placement, and authors' rationales.}
As shown in Table~\ref{table:3.3}, textbooks originating in the twentieth century largely reflect a pre--quantum-information curriculum that omits modern foundational and interpretational topics, whereas contemporary texts and revised editions increasingly integrate them—with Ballentine '15$^2$ explicitly citing thirty years of experimental and theoretical advances in quantum foundations as a primary motivation for new editions. Structural prominence varies sharply: some texts concentrate entanglement, measurement, and foundations into an intensive early chapter (or across multiple dedicated chapters), whereas others deliberately restrict philosophical and interpretational discussions to late chapters or appendices, arguing in prefaces that students must master the ``orthodox'' formalism and its applications before engaging with foundational controversies.

\paragraph{Entanglement and nonlocality.}
Included in almost all post-2000 corpus texts, entanglement has shifted from a late topic tied exclusively to the EPR paradox and Bell's theorem to a central, recurring theme—especially in QIST-aligned textbooks. Across the corpus, entanglement is contextualized along four distinct lines: (i) historically and foundational-theoretically via EPR and Bell nonlocality; (ii) through quantum-information applications (e.g., teleportation and computing); (iii) mathematically via tensor products, superposition in composite Hilbert spaces, and reduced density operators; or (iv) physically through many-body systems, measurement theory, and decoherence. Nonlocality is similarly framed either through Bell inequalities, quantum interference, or broader accounts of tunneling and quantum potentials.

\paragraph{Interpretations, the measurement problem, and decoherence.}
Textbook chapter structures situate these three topics within varied organizing contexts: justifying the theory, comparing alternative interpretations, analyzing physical measurement processes, or describing open-system dynamics. Coverage of interpretations ranges from implicit reliance on the Copenhagen view to dedicated opening or closing chapters comparing competing frameworks. Similarly, the measurement problem is framed either within quantum foundations, alongside formal measurement postulates, or implicitly via thought experiments such as Schrödinger's cat. Finally, while some texts discuss the measurement problem without decoherence, many contemporary texts link the two directly—framing decoherence both as a resolution to practical aspects of the measurement problem and as an essential tool for modeling environmental interactions, open quantum systems, and the quantum-to-classical transition.

\section{Pedagogical pathways}
\label{sec:res_RQ2}
Our topic-level analyses in Sec.~\ref{sec:res_RQ1} demonstrate that the pedagogical role of a topic is not intrinsic to the topic itself, but depends on its sequencing and cross-dimensional contextualization---challenging any simple separation of topics between a fixed ``canonical core'' and later foundational or modern additions. Building on these findings, we now synthesize how mathematical tools, phenomena and experiments, and theoretical concepts are coordinated across the curriculum to characterize textbooks' overarching pedagogical pathways according to their \textit{primary entry point} (Sec.~\ref{sec:res_4.1}, addressing RQ2.1--RQ2.2) and their overall \textit{pedagogical orientation} (Sec.~\ref{sec:res_4.2}, addressing RQ2.3), culminating in a practical roadmap for textbook selection and curriculum design (Sec.~\ref{sec:res_4.3}).

In the broader textbook literature, authors' stated rationales reveal competing pedagogical priorities in how entry points are chosen and structured. Some authors favor a historical route in which QM emerges from experimental and conceptual crises \cite{Hameka_2004}, whereas others open with formal postulates for their logical coherence and capacity to foreground distinctively quantum phenomena early \cite{bes_2012, atkins_2011}, and still others prioritize building physical intuition through qualitative explanations and classical analogies before introducing formal calculations \cite{scherrer_2024}. Consequently, an entry point (RQ2.1, RQ2.2) is characterized not merely by its initial physical system or mathematical tool—as labels like ''wave-first'' and ''spin-first'' imply—but by how diverse topic-level pedagogical choices (RQ1) and authors' rationales for introducing QM are coordinated to motivate the theory.

Beyond the entry point, a textbook's overall pedagogical orientation captures the broader priorities sustained throughout the work, reflecting its overarching purpose and intended audience. As Basdevant '23$^3$ notes, QM is an ``ideal subject'' because it can be approached from multiple perspectives, including physics, mathematics, technology, philosophy, and culture. As reflected in authors' stated rationales in the broader textbook literature, a textbook may foreground fundamental theoretical physics \cite{Weinberg_2015}, conceptual clarity and mathematical structure \cite{schwindt_2016}, or the historical and conceptual development of QM \cite{peebles_2020}. Although these broad orientations are not mutually exclusive, their relative prominence—reflected in how topic-level pedagogical choices (RQ1) are sustained throughout the text—reveals the pedagogical orientation (RQ2.3).

\subsection{Entry point}
\label{sec:res_4.1}

\subsubsection{A multidimensional framework of entry-point archetypes}

Our topic-level analyses (Sec.~\ref{sec:res_RQ1}) show that a textbook's mathematical, physical, and conceptual entry points need not coincide: conceptual illustrations---notably single-photon setups paired with Dirac notation---frequently precede worked calculations using wavefunctions or spin matrices (Secs.~\ref{sec:res_1}, \ref{sec:res_2}), while foundational postulates may structure the opening deductively or consolidate concepts developed through earlier examples (Sec.~\ref{sec: res_3.1}).

Moreover, authors' rationales justify their opening sequences through diverse priorities---mathematical accessibility, conceptual generality, curricular progression, and preparation for modern applications---that do not uniquely determine a single physical system or mathematical formalism. Similar goals can motivate different topic configurations, and several authors explicitly reject the wave-spin binary, seeking to circumvent perceived limitations of both approaches.

Characterizing an entry point therefore requires looking beyond isolated topic selection and sequencing to examine \textit{contextualization} across the three dimensions---specifically, how initial \textit{mathematical tools} represent states and dynamics, how \textit{phenomena and experiments} (whether conceptual illustrations or worked models) physically ground that formalism, and how \textit{theoretical concepts} link the two to explain quantum behavior and organize the emerging rules.

Synthesizing the topic-level findings from Sec.~\ref{sec:res_RQ1} alongside PER literature, we characterize five entry-point archetypes (Table~\ref{table:qm_pedagogy}). Addressing RQ2.1, we retain \textbf{``wave-first''} (matter-wave/history) and \textbf{``spin-first''} (two-state/spin-$1/2$), grounding them in their commonly associated topics within individual dimensions. Addressing RQ2.2, we add three archetypes reflecting alternative approaches in the corpus and broader literature: \textbf{``path-first''} (interference/path amplitudes), \textbf{``qubit-first''} (quantum information/technology), and \textbf{``axiom-first''} (postulational/formal). The table also summarizes their potential pedagogical advantages and challenges drawn from authors' rationales and PER studies.

These labels serve as concise names for recurring multidimensional configurations rather than rigid categories defined by necessary conditions. The dimensions need not carry equal weight: an archetype may be distinguished primarily by its mathematical approach, motivating phenomena, or theoretical emphasis. Thus, substantial overlap in one or two dimensions need not imply the same archetype: spin-first and qubit-first openings both use two-dimensional state spaces and often the same physical realizations (e.g., photon polarization), but qubit-first openings are distinguished by their theoretical emphasis---explicitly adopting qubit terminology and an information-theoretic framing from the outset. Conversely, a single topic can support multiple archetypes depending on how it is framed: Dirac notation can anchor spin-, qubit-, or axiom-first openings; Stern--Gerlach experiments can introduce spin measurement, frame two-state systems as qubits, or illustrate measurement axioms; and historical experiments often motivate wave-first openings without being required by or exclusive to them.

\begin{widetext}
\begin{table}[H]
    \centering
    \parbox{\textwidth}{
      \caption{Entry-point archetypes and their commonly associated mathematical tools, phenomena/experiments, theoretical concepts, and potential pedagogical advantages and challenges (RQ2.1, RQ2.2). The listed features describe recurring configurations and characterizations summarized from textbooks and PER literature, rather than required checklists; the presence, timing, and relative emphasis of individual features vary across textbooks. \\}
      \label{table:qm_pedagogy}
    }    
    \renewcommand{\arraystretch}{1.2} 
    \setlength{\tabcolsep}{5pt}
    \footnotesize
    \resizebox{1.0\textwidth}{!}{
    \begin{tabular}
    {p{0.12\textwidth}|p{0.15\textwidth}p{0.17\textwidth}p{0.15\textwidth} |p{0.20\textwidth}p{0.21\textwidth}}
        \toprule
        \textbf{Entry-point archetype} & \textbf{Mathematical tools} & \textbf{Phenomena/ \newline experiments} & \textbf{Theoretical \newline concepts} & \textbf{Potential pedagogical advantages} & \textbf{Potential pedagogical challenges} \\
        \midrule

        {Matter-wave/ History} \newline \newline (``wave-first'') & 
        Wave mechanics, partial differential equations, calculus &
        Blackbody radiation, photoelectric effect, atomic spectra; free motion and confinement & 
        Wave--particle duality, quantized energy, Old Quantum Theory & 
        Classical--quantum transition [Townsend '12$^2$, Lvovsky '18$^1$], tracks rational theory development \cite{atkins_2011}  & 
        Quantum concepts hidden behind heavy math [Binney '14$^1$, Kok '23$^2$, Larkoski '23$^1$, Lvovsky '18$^1$], potential classical misconceptions \cite{Fischler_1992} \\ \hline
        
        {Two-State/ Spin-1/2} \newline \newline (``spin-first'') & 
        Matrix mechanics, Dirac notation, finite-dimensional linear algebra & 
        Stern--Gerlach, photon polarization, ammonia molecule, double-well & 
        Quantum measurement, incompatible observables, superposition & 
        Isolates quantum concepts early, avoids calculus [McIntyre '22$^1$, Townsend '12$^2$] & 
        Possible struggle transitioning from discrete to continuum \cite{Solorio_2025} \\ \hline 
        
        {Interference/ Path amplitudes} \newline \newline (``path-first'') & 
        Complex probability amplitudes, phasors, elementary linear algebra; path sums or integrals & 
        Double-slit experiment, Mach--Zehnder interferometers & 
        Superposition, probability amplitudes, indistinguishability &
        Phenomenon-based \cite{Marshman_2017}, simple math \cite{Taylor_1998}, classical--quantum connection \cite{Taylor_1998}, potential connection to advanced physics \cite{Taylor_1998} &  
        Awkward in analyzing certain systems \cite{Taylor_1998}  \\ \hline  
        
        {Q. Information/ \newline Q. Technology }  \newline \newline (``qubit-first'') & 
        Dirac notation, finite-dimensional linear algebra, tensor products, density operator & 
        Modern quantum experiments, Bell test, polarization, selected atomic levels, spins, or idealized information-processing tasks & 
        Superposition, entanglement, nonlocality, entropy, two-state preparation, measurement, quantum information & 
        Close connection to modern laboratories and applications [Lvovsky '18$^1$], initially simple math \cite{carlini_2026}  & 
        Bridging abstract formalism and physical interpretation, potentially requiring supporting laboratory activities \cite{carlini_2026}
         \\ \hline
        
        {Postulated/ Formal} \newline \newline (``axiom-first'') & 
        Hilbert space, operator theory, eigenvalue problems, functional analysis in advanced versions &
        Generalized physical system; harmonic oscillator \cite{Bohm_1986}; single-photon/quantum optical experiments & 
        Foundational statements of QM (Tables~\ref{table:3.1}, \ref{table:3.2}) & 
        A unified framework, a mathematically clean foundation [Pade '18$^2$] \cite{atkins_2011} & 
        Cognitive barrier [Shankar '94$^2$], unmotivating without empirical motivation or phenomena-based scaffolding [Pade '18$^2$] \\ 
        \bottomrule
    \end{tabular}
    }
\end{table}
\end{widetext}

\subsubsection{Analytical utility: Capturing variation, hybridization, and cross-level comparison}
Applying the framework to each text's initial development of QM yields the entry-point characterizations reported in the second column of Table~\ref{table:4} for our 36-textbook corpus (RQ2.2). The following applications show how the framework goes beyond assigning labels to reveal variation within archetypes, map combinations across them, and support comparisons of curricular design and pedagogical trade-offs. The five archetypes do not encompass every opening configuration in the corpus. Where assigning any archetype is inadequate, we retain descriptive characterizations, such as the heuristic opening in Feynman '11 ('65) and the interpretation- and history-oriented opening in Cetto '26$^1$. Nevertheless, these cases remain analyzable across the three dimensions.

\paragraph{Resolving variation within archetypes.}
A primary strength of this three-dimensional framework is its ability to  resolve important pedagogical differences within a single archetype that a broad label alone obscures. Griffiths '18$^3$, Basdevant '23$^3$, Messiah '95$^2$, and Dick '20$^3$, for example, are all characterized primarily as \textit{wave-first}, yet they show markedly different balances across the three dimensions: Messiah '95$^2$ and Basdevant '23$^3$ develop extensive experimental and historical motivation before or alongside the wave function; Dick '20$^3$ connects its opening wave-mechanical treatment to modern physical applications while advancing rapidly toward formal mathematical tools. By contrast, Griffiths '18$^3$ begins directly with the time-dependent Schrödinger equation and its statistical interpretation, devoting little opening space to experimental motivation.
Comparable internal variations also appear within other predominant archetypes, such as \textit{axiom-first} introductions that differ substantially in how they scaffold mathematical tools.

\paragraph{Deconstructing hybrid entry points via concurrent and sequential notation.}
The framework also captures how multiple archetypes can be combined. Table~\ref{table:4} uses two notations: ``+'' indicates \textbf{concurrent integration}, in which multiple archetypes are interwoven within the initial development of QM; a comma indicates \textbf{sequential combination}, in which one approach does not receive sustained development, but is followed by another. 

This notation reveals structural distinctions that binary labels obscure. Most notably, while ``spin-first'' is often treated in the literature as a broad category, applying the framework shows that Townsend '12$^2$ is the only textbook coded as a standalone \textit{spin-first} entry point; all other spin-based introductions concurrently or sequentially integrate other approaches.

Similarly, the notation captures sequential scaffolding—such as historical wave openings consolidated by postulates (\textit{wave, axiom} in Liboff '03$^4$) or optical path interference preceding wave mechanics (\textit{path, wave} in Dunningham '18$^2$ and Zwiebach '22$^1$)—as well as compound trajectories like Cohen-T.\ '20$^2$ (\textit{wave $+$ path, axiom}). 

Analytically, this notation provides a basis for examining how combinations may address the potential limitations of individual archetypes: for instance, combining wave-first with path- or axiom-first approach to foreground measurement early, or grounding an abstract axiom- or qubit-first opening in concrete optical interference.

\paragraph{Examining curricular alignment across instructional levels.}
Comparing \textit{introductory} and \textit{advanced} texts (unshaded and shaded rows in Table~\ref{table:4}) shows how the framework can help identify potential points of curricular alignment and discontinuity. Openings of advanced texts may reorganize familiar topics rather than simply review them: notably, 10 of the 14 advanced texts incorporate the \textit{axiom-first} archetype into their entry point. For students whose prior course began with and remained strongly oriented toward wave mechanics, for example, the predominantly axiomatic opening of Ballentine '15$^2$ or the integrated \textit{spin $+$ axiom $+$ wave} opening of Sakurai '21$^3$ may introduce a different relationship between mathematical representations, physical systems, and theoretical concepts. The potential transition therefore concerns not only familiarity with individual topics, but also how QM is organized from the outset. Used alongside information about students' prior coursework, the framework offers a structured basis for identifying connections that can be made explicit and shifts in framing that may warrant additional scaffolding.

\paragraph{Bridging established textbooks and emerging QIST curricula.}
Finally, the framework situates established textbook approaches in relation to emerging curricular designs across educational levels. In our corpus, neither \textit{path-first} nor \textit{qubit-first} appears as a standalone entry point; both occur within combinations with other archetypes. This pattern contrasts with emerging pre-university learning designs that use qubits as the primary organizing basis for introducing QM \cite{carlini_2026}. Related studies retain the ``spin-first'' label while foregrounding qubits and an information-theoretic perspective \cite{Albert_2025}. By distinguishing approaches through the contextualization of topics across the three dimensions, the framework enables comparison beyond the labels adopted by individual authors.
Broader efforts to introduce QIST before university \cite{Sun_2024} raise questions about how qubit-based learning might support subsequent formal study of QM. The framework provides a shared analytical vocabulary for examining this curricular alignment.

\subsection{Overall pedagogical orientations}
\label{sec:res_4.2}
Identifying a textbook's entry point using our three-dimensional framework characterizes its initial illustration of QM, but does not necessarily capture the priorities that govern its subsequent development. We therefore extend the framework across the full textbook trajectory to profile each work's overall \textit{pedagogical orientation} (RQ2.3), with titles and prefaces providing complementary evidence.

\subsubsection{A multidimensional framework of pedagogical orientations}
\paragraph{Orientation dimensions and qualitative criteria.}
Mirroring our three analytical dimensions, we profile textbooks along three non-exclusive orientations: (i) \textbf{mathematical orientation}, concerning the extent to which systematic mathematical development and justification serve as an independent instructional objective rather than solely as a tool for physical applications; (ii) \textbf{experiment/application orientation}, concerning the organizing role of empirical phenomena, laboratory realizations, and technological applications; and (iii) \textbf{theoretical-physics orientation}, concerning the development of general physical theory and its connections to broader theoretical frameworks (such as statistical mechanics, condensed-matter theory, or QFT). Within the experiment/application orientation, we also record a ``modernity'' sub-indicator to distinguish whether the foregrounded contexts center primarily on historical experiments motivating the emergence of QM (low, $+$), established twentieth-century quantum technologies such as lasers (moderate, $++$), or the sustained integration of contemporary experiments and applications (high, $+++$).

As summarized in Table~\ref{table:criteria}, each orientation is rated qualitatively as low ($+$), moderate ($++$), or high ($+++$) depending on whether that dimension plays a supplementary role, a recurring supporting role, or a sustained organizing role across the exposition. 

\begin{widetext}
\begin{table}[H]
\centering
\parbox{\textwidth}{
      \caption{Qualitative criteria for profiling the overall pedagogical orientation of QM textbooks across the three dimensions (RQ2.3). Ratings describe the structural role of each orientation across the textbook, not its technical difficulty, rigor, or instructional quality. Because the dimensions are non-exclusive, a textbook may receive a high rating in more than one dimension. Named textbooks illustrate applications of the criteria.}
      \label{table:criteria}
    }   
\renewcommand{\arraystretch}{1.2}
\setlength{\tabcolsep}{5pt}
\footnotesize
\resizebox{1.0\textwidth}{!}{
\begin{tabular}{p{0.09\textwidth}|p{0.29\textwidth}p{0.29\textwidth}p{0.29\textwidth}}
\toprule
\textbf{Rating} & \textbf{Mathematical orientation} & \textbf{Exp./App. orientation} & \textbf{Theoretical-Physics orientation} \\
\midrule
Low \newline ($+$)
& Mathematical tools are developed chiefly for immediate physical use, with limited independent organization of mathematical material. [Townsend '12$^2$]
& Exp./App. are absent, occasional, or supplementary, with little influence on the overall sequence of development. [Wan '19$^1$] 
& Broader theoretical connections are limited or supplementary and do not substantially organize the exposition. \\ \hline

Moderate \newline ($++$) 
& Mathematical structures receive systematic treatment in dedicated sections or chapters, but do not provide a sustained organizing thread across the text.
& Exp./App. recur and support substantial portions of the discussion, but do not provide a sustained organizing thread across the text. [Townsend '12$^2$] 
& General theoretical structures and connections to further areas of physics receive substantial treatment, but their organizing role is mainly confined to specific chapters. [Townsend '12$^2$] \\ \hline

High \newline ($+++$) 
& The systematic development and justification of mathematical structures provide a sustained organizing thread, shaping topic selection, sequencing, and exposition of QM. [Wan '19$^1$]
& Exp./App. provide a sustained organizing thread, shaping topic selection, sequencing, and explanation across the textbook. [Simon '25$^1$] 
& The development of general physical theory and its connections to broader theoretical frameworks provide a sustained organizing thread across the text. [Dick '20$^3$] \\

\bottomrule
\end{tabular}
}
\end{table}
\end{widetext}

\paragraph{Authors' stated priorities and pedagogical trade-offs.}
While a textbook's title sometimes signals its intended orientation---such as foregrounding theoretical principles (Dirac '67$^4$, Shankar '94$^2$), mathematical foundations (Larkoski '23$^1$), or experimental and technological contexts (Haroche '06$^1$, Simon '25$^1$)---our classifications are grounded in the structural organization of content across the book, informed by authors' explicit rationales in prefaces. These prefaces reveal how different goals and audiences drive divergent orientation profiles:
\begin{itemize}
    \item \textit{Foregrounding theoretical structure:} Sakurai '21$^3$ emphasizes abstract states, operators, and representations to present wave mechanics as just one realization of quantum dynamics and to clarify how classical physics emerges as an approximation. Similarly, Năstase '23$^1$ targets readers for whom constructing the theoretical formalism and reaching advanced topics beyond the standard curriculum are preferred over extended introductory motivation.  
    \item \textit{Foregrounding experiments and applications:} Conversely, Commins '14$^1$ explains that his topic selection---shaped in part by his own experimental background---prioritizes physical phenomena and applications suited to students entering experimental physics or adjacent scientific fields who may never study particle physics or QFT. 
    \item \textit{Bounding mathematics:} Prefaces also explicitly justify low ratings in a given dimension: Kok '23$^2$, for instance, states that mathematical techniques are introduced strictly on a ``need-to-know'' basis to avoid overshadowing the physical discussion.
\end{itemize}

\subsubsection{Analytical utility: Profiling curricular balance, multi-orientation designs, and modernity}
Applying the criteria in Table~\ref{table:criteria} to the full curricular trajectory of each text yields the pedagogical-orientation profiles reported in the third column of Table~\ref{table:4} for our 36-textbook corpus (RQ2.3). 
Beyond characterizing the sampled texts, the following applications demonstrate how this profiling framework can be used for broader textbook and curricular comparisons---distinguishing structural emphasis from technical difficulty, identifying strategies for connecting multiple orientations, and tracking engagement with modern experimental and application contexts beyond publication date and QIST.

\paragraph{Distinguishing dominant curricular emphasis from technical difficulty.} 
Across the corpus, 21 textbooks exhibit a single dominant orientation across the three dimensions: one is led primarily by a \textbf{mathematical orientation} (Wan '19$^1$, which develops extensive Hilbert-space mathematics before systematically presenting the physical framework, while concentrating concrete applications later); six are led by an \textbf{experiment/application orientation} (such as Simon '25$^1$, where quantum computation, communication, and measurement organize the exposition, yielding $+$$+$$+$ experiment/application with high modernity); and 14 are led by a \textbf{theoretical-physics orientation} (such as Dick '20$^3$, which sustains connections to condensed-matter and photon theory throughout). 

Crucially, these profiles capture the \textit{structural role} of each dimension rather than technical difficulty: for example, both Townsend '12$^2$ and Simon '25$^1$ receive a low ($+$) mathematical-orientation rating despite substantial differences in mathematical complexity, because in both texts mathematical tools are introduced primarily in service of physical and technological discussions. This distinction separates two questions that are easily conflated in curricular comparisons: how technically demanding the mathematical or theoretical material is, and whether its development serves to organize the curriculum or primarily to support other instructional goals.

\paragraph{Mapping multi-orientation designs across the curriculum.}
While the 21 texts above are coded with a unique highest-rated orientation, the remaining 15 textbooks demonstrate that many textbooks adopt multi-orientation designs (in which two or all three orientations receive equally high ratings):

\begin{itemize}
    \item \textit{Experiment/Application $+$ Theoretical Physics (6 texts):} Exemplified by Townsend '12$^2$, which balances moderate theoretical and experimental orientation against a low mathematical-orientation rating, and Levy '26, which weaves a recurring theoretical theme---the discrete-to-continuous transition across state spaces, spatial and temporal resolution, and scale transformations---together with sustained quantum-information applications, progressing from introductory qubits to logical qubits and quantum error correction.
    \item \textit{Theoretical Physics $+$ Mathematics (4 texts):} Exemplified by Larkoski '23$^1$, which combines rigorous, sustained mathematical development with the construction of the quantum-theoretical framework while assigning a limited organizing role to experimental applications.
    \item \textit{All Three Orientations (5 texts):} Exemplified by Freericks '26$^1$, which systematically integrates conceptual and experimental optical setups, representation-independent operator mathematics, and theoretical development across the textbook---realizing the preface's stated progression from conceptual ideas to the requisite mathematics and formal framework, and followed by applications to experimental situations.
\end{itemize}
Beyond classifying individual books, these profiles identify examples of how multiple curricular goals can be connected, whether through an overarching theoretical theme (Levy '26) or a staged progression (Freericks '26$^1$). For curriculum developers, these examples offer strategies that can be adapted to local instructional goals. For PER researchers, they motivate investigation of whether and how these curricular connections help students relate mathematical, theoretical, and experimental aspects of QM.

\paragraph{Decoupling experiment/application modernity from publication era.}
Finally, our ``modernity'' sub-indicator identifies engagement with modern experimental and application contexts beyond QIST, including cosmological applications such as the cosmic microwave background in Dick '20$^3$. Such contexts appear in both recent textbooks and several earlier volumes, indicating that publication date alone is an insufficient proxy for engagement with modern physics. For educators updating quantum courses and researchers studying curricular reform, this finding broadens the resources and contexts worth considering: established textbooks may offer relevant material, and modernization need not be equated exclusively with quantum information and computing.

\begin{table*}[h!]
\caption{Identified pedagogical pathways of the selected texts:
primary entry points, pedagogical orientations (mathematical orientation, experiment/application orientation (with the ``modernity'' sub-indicator), theoretical physics orientation), and scope and thematic emphasis of each textbook. Green shading marks entry points classified under a single archetype. Blue shading marks orientations rated high (+++), including the ``modernity'' sub-indicator of experiment/application orientation where applicable. Unlike the other tables, texts are grouped here by their
dominant orientation(s); within each group, the ordering convention
used elsewhere is retained.}
\label{table:4}
\begin{threeparttable}
\centering
\footnotesize
\renewcommand{\arraystretch}{1.15} 
\resizebox{1.0\textwidth}{!}{%
\begin{tabular}{|p{0.16\textwidth}|p{0.18\textwidth}|c|c|c|c|p{0.34\textwidth}|}
\toprule
\mbox{} & \mbox{} & \mbox{} & \multicolumn{2}{c|}{\textbf{Exp./App.}} & \textbf{Theo.} & \mbox{} \\ 
\cline{4-5} 
Text ID & \textbf{Primary entry point} & \textbf{Math.} & \scriptsize\textit{gen.} & \scriptsize\textit{modern.} & \textbf{phy.} & Scope and thematic emphasis \\ 
\hline\midrule

\cellcolor{gray!45}Wan '19$^1$ & \cellcolor{green!25}axiom & \cellcolor{blue!25}+++ & + & \scriptsize + & + & Rigorous math. \& foundation \\ 
\midrule

\textbf{Feynman '11 ('65)} & heuristic & ++ & \cellcolor{blue!25}+++ & \scriptsize ++ & ++ & Phenomenological, physical insight \\ 
\textbf{Lvovsky '18$^1$} & axiom + qubit + path & + & ++ & \cellcolor{blue!10}\scriptsize +++ & + & QIST-preparation \\ 
\textbf{Kok '23$^2$} & path, spin & + & ++ & \cellcolor{blue!10}\scriptsize +++ & + & QIST-preparation \\ 
\cellcolor{gray!25}Haroche '06$^1$ & axiom + qubit & ++ & \cellcolor{blue!25}+++ & \cellcolor{blue!10}\scriptsize +++ & ++ & QIST preparation; microwave CQED-focused \\ 
\cellcolor{gray!25}Beck '12$^1$ & path + axiom & + & \cellcolor{blue!25}+++ & \cellcolor{blue!10}\scriptsize +++ & + & Applied optics, hands-on lab-integrated \\ 
\cellcolor{gray!25}Simon '25$^1$ & qubit, path & + & \cellcolor{blue!25}+++ & \cellcolor{blue!10}\scriptsize +++ & + & QIST-preparation \\ 
\midrule

Dirac '67$^4$ & path, axiom & ++ & + & \scriptsize + & \cellcolor{blue!25}+++ & Foundational, abstract \\ 
Shankar '94$^2$ & wave, axiom & ++ & + & \scriptsize + & \cellcolor{blue!25}+++ & Broad coverage \\ 
Messiah '95$^2$ & \cellcolor{green!25}wave & ++ & + & \scriptsize + & \cellcolor{blue!25}+++ & Systematic, detailed, broad reference \\ 
Gottfried '03$^2$ & path + wave, axiom & ++ & ++ & \scriptsize ++ & \cellcolor{blue!25}+++ & Conceptual foundation; QFT-preparation \\ 
Liboff '03$^4$ & wave, axiom & ++ & + & \scriptsize + & \cellcolor{blue!25}+++ & Solid state/CM preparation \\ 
Griffiths '18$^3$ & \cellcolor{green!25}wave & + & + & \scriptsize + & ++ & Elementary \\ 
Banks '19$^1$ & spin + axiom & ++ & ++ & \scriptsize ++ & \cellcolor{blue!25}+++ & Operator \& symmetry-driven; CM preparation, QFT-adjacent \\ 
Cohen-T. '20$^2$ & wave + path, axiom & ++ & ++ & \scriptsize ++ & \cellcolor{blue!25}+++ & Encyclopedic; AMO/atomic physics-preparation \\ 
\cellcolor{gray!25}Dick '20$^3$ & \cellcolor{green!25}wave & ++ & + & \scriptsize ++ & \cellcolor{blue!25}+++ & CM \& QFT-preparation; rigorous math \\ 
\cellcolor{gray!45}Landau '77$^3$ & axiom, wave & ++ & + & \scriptsize + & \cellcolor{blue!25}+++ & Compact, analytical, physical insight \\ 
\cellcolor{gray!45}\textbf{Jackson '24 ('89)} & wave, spin + axiom & ++ & + & \scriptsize + & \cellcolor{blue!25}+++ & Analytical; QFT-preparation \\ 
\cellcolor{gray!45}Ballentine '15$^2$ & \cellcolor{green!25}axiom & ++ & + & \scriptsize ++ & \cellcolor{blue!25}+++ & Symmetry-driven; QIST-integrated; rigorous math. \\ 
\cellcolor{gray!45}Sakurai '21$^3$ & spin + axiom + wave & ++ & + & \scriptsize + & \cellcolor{blue!25}+++ & Formalism/symmetry-oriented \\ 
\cellcolor{gray!45}Martin '26$^1$ & wave + spin & ++ & + & \scriptsize ++ & \cellcolor{blue!25}+++ & Comprehensive, strong atomic/AMO \\ 
\midrule

Binney '14$^1$ & axiom + path, wave & ++ & + & \scriptsize ++ & ++ & Atomic physics/AMO preparation \\ 
Larkoski '23$^1$ & \cellcolor{green!25}axiom & ++ & + & \scriptsize ++ & ++ & Operator-centered \\ 
Cetto '26$^1$ & interpretation, history & ++ & + & \scriptsize ++ & ++ & History-integrated \\ 
\cellcolor{gray!45}Năstase '23$^1$ & axiom, spin + qubit & \cellcolor{blue!25}+++ & ++ & \scriptsize ++ & \cellcolor{blue!25}+++ & Comprehensive \\ 
\midrule

Dunningham '18$^2$ & path, wave & + & ++ & \scriptsize ++ & ++ & Broad introductory coverage \\ 
McIntyre '22$^1$ & spin + axiom & + & ++ & \scriptsize ++ & ++ & \mbox{} \\ 
\textbf{Styer '24} & spin + path + qubit & + & ++ & \scriptsize ++ & ++ & \mbox{} \\ 
\cellcolor{gray!45}Townsend '12$^2$ & \cellcolor{green!25}spin & + & ++ & \scriptsize ++ & ++ & \mbox{} \\ 
\cellcolor{gray!45}\textbf{Levy '26} & spin + axiom, qubit & ++ & \cellcolor{blue!25}+++ & \cellcolor{blue!10}\scriptsize +++ & \cellcolor{blue!25}+++ & QIST-integrated; many-body quantum information \\ 
\cellcolor{gray!45}Commins '14$^1$ & axiom, wave & ++ & \cellcolor{blue!25}+++ & \cellcolor{blue!10}\scriptsize +++ & \cellcolor{blue!25}+++ & Experimental particle phenomenology \\ 
\midrule

Auletta '09$^1$ & path + axiom & ++ & ++ & \cellcolor{blue!10}\scriptsize +++ & ++ & Modern foundation \& quantum information \\ 
\textbf{Pade '18$^2$} & wave, spin & ++ & ++ & \cellcolor{blue!10}\scriptsize +++ & ++ & QIST-integrated; Hybrid analytical/algebraic \\ 
Zwiebach '22$^1$ & path, wave & ++ & ++ & \cellcolor{blue!10}\scriptsize +++ & ++ & \mbox{} \\ 
\textbf{Basdevant '23$^3$} & \cellcolor{green!25}wave & ++ & ++ & \cellcolor{blue!10}\scriptsize +++ & ++ & Broad application, e.g. astronomy \\ 
Freericks '26$^1$ & spin, path & ++ & ++ & \cellcolor{blue!10}\scriptsize +++ & ++ & Operator-centered; QIST-integrated; strong quantum optics/AMO/CM \\ 
\bottomrule

\end{tabular}}

 \begin{tablenotes}
\item \textit{Abbreviations:} \textit{gen.} = general experimental/applied orientation; AMO = atomic, molecular, and optical physics; CM = condensed matter physics; CQED = cavity quantum electrodynamics; QFT = quantum field theory; QIST = quantum information science and technology.
\item A blank entry indicates that no specific thematic emphasis was identified.
 \end{tablenotes}
 
\end{threeparttable}

\end{table*}

\subsection{A Practical Roadmap for Textbook Selection and Curriculum Design}
\label{sec:res_4.3}
Although textbooks consolidate established knowledge and are often read as authoritative accounts of a discipline, our topic-level analyses (Sec.~\ref{sec:res_RQ1}) and pathway syntheses (Secs.~\ref{sec:res_4.1}--\ref{sec:res_4.2}) underscore that they remain selective pedagogical constructions. 
Notably, combining structural analysis with authors' rationales shows that even shared broad goals---such as conceptual clarity and pedagogical accessibility---can motivate opposite structural arrangements: they are invoked both to justify an early postulate-based presentation and to justify deferring axioms until after physical examples (Sec.~\ref{sec: res_3.1}), just as they are used to motivate either building gradually from familiar classical physics or confronting students immediately with distinctively quantum two-state behavior (Secs.~\ref{sec:res_1}--\ref{sec:res_2}). Because each configuration carries distinct advantages and challenges, no single textbook profile is universally optimal.

Practically, our framework supports textbook selection and curriculum design by making these underlying structural choices explicit---surfacing curricular priorities and potential gaps that instructors and learners might otherwise encounter only after working through a substantial portion of a book. Together with closer examination of candidate texts, the framework helps evaluate key curricular aspects of alignment with \textit{learners' preparation} (prior mathematics, previous QM exposure, and disciplinary background; cf.\ Table~\ref{table:books}), \textit{instructional goals} (desired orientation, target applications, and downstream coursework), and \textit{course constraints} (available time and required coverage).
Using our empirical tables as comparative starting points, the following checklist translates the framework into guiding questions for curricular review and closer examination of candidate texts:
\begin{itemize}
    \item \textbf{Establish the intended overall orientation} (Tables~\ref{table:criteria} and~\ref{table:4}): Determine the desired balance among mathematics, experiment/application, and theoretical physics across the full course rather than inferring priorities from an opening chapter alone. The thematic descriptors in Table~\ref{table:4} further help match specific disciplinary foci (e.g., many-body physics, field theory, or quantum optics).

    \item \textbf{Evaluate the entry point and opening sequence} (Tables~\ref{table:qm_pedagogy} and~\ref{table:4}): Examine how initial mathematical tools, motivating phenomena, and theoretical concepts are selected, sequenced, and contextualized. Where archetypes are combined, consider what concurrent integration (``$+$'') or sequential combination (``$,$'') assumes about students' preparation and what connections must be made explicit.

    \item \textbf{Map the mathematical trajectory} (Table~\ref{table:1}): Check when core representations (wave mechanics, matrix mechanics, Dirac notation) and extended formalisms (path integrals, density operators, second quantization) are introduced, what prerequisites they require, and how the text supports translations across representations.

    \item \textbf{Check the role and framing of physical examples} (Table~\ref{table:2}): Distinguish systems used for conceptual illustration from those receiving worked mathematical treatment, and assess the roles of finite- and infinite-dimensional state spaces and their concrete physical realizations.

    \item \textbf{Examine the organization of foundational statements} (Tables~\ref{table:3.1} and~\ref{table:3.2}): Consider whether postulates and principles organize the opening deductively, consolidate earlier examples, or recur in progressively more general formulations, and identify where links to physical or mathematical preparation require extra scaffolding.

    \item \textbf{Determine the treatment of foundations and interpretations} (Table~\ref{table:3.3}): Examine how the text distinguishes and relates operational measurement rules, the measurement problem, decoherence, and competing interpretations. Decide whether brief contextual discussion, sustained analysis, or supplementary readings best fit the instructional goals.

    \item \textbf{Assess contemporary applications and QIST integration} (Tables~\ref{table:2},~\ref{table:3.3}, and~\ref{table:4}): Identify relevant experimental platforms and applications within and beyond QIST without treating publication date as a proxy for modernity. Where quantum information is a priority, examine the placement, depth, and contextualization of qubits, composite systems, entanglement, density operators, no-cloning, and decoherence.
\end{itemize}
Crucially, our framework does not imply that a single textbook must fulfill every instructional aim; instead, it can guide the combination of complementary resources—using one textbook for its overarching trajectory while drawing on another for alternative physical illustrations, representational bridges, or extended treatments of foundations and modern applications.

\section{Conclusion and Outlook}
\label{sec:conclusion}
By comparatively analyzing 36 university-level QM textbooks, this study provides a broad picture of the patterns and variations in pedagogical choices that shape the contemporary landscape of university quantum education---demonstrating that pedagogical diversity in quantum education extends far beyond topic coverage and the conventional ``wave-first''/``spin-first'' binary. Across mathematical tools, phenomena and experiments, and theoretical concepts, textbooks differ not only in which topics they select and when they sequence them, but critically in how they contextualize and connect them (RQ1)---choices that we synthesize through five entry-point archetypes (and their concurrent or sequential combinations) as well as broader pedagogical orientations across the curriculum (RQ2). Authors justify these varied designs through recurring, differently balanced rationales concerning mathematical accessibility, theoretical generality, physical motivation, and preparation for modern applications. By making these structural choices explicit, our framework and practical roadmap support curriculum design and the purposeful combination of textbook resources without prescribing a single universally optimal pathway: the central question is not simply which label a textbook or a curriculum design carries, but how its multidimensional organization connects learners' preparation to the intended goals of instruction.

Several methodological boundaries of this study point directly to opportunities for extending the framework. Because our corpus was purposively selected to maximize structural variation across English-language textbooks, reported counts describe the analyzed sample rather than market prevalence or historical trends; applying the framework to a broader, multilingual corpus would test its transferability across national curricular traditions. At the topic level (RQ1), recording chapter-level placements and recurring contextualization patterns captures macro-structural design rather than fine-grained instructional time, worked-example solutions, end-of-chapter exercises, local transitions among verbal, mathematical, and graphical representations, or external digital supplements. Likewise, textbook prefaces capture only rationales that authors state explicitly. At the pathway level (RQ2), while chapter-level topic locations were double-coded on a subset of texts, entry-point classifications and orientation ratings represent holistic qualitative syntheses by one author. Because curricular boundaries and emphases are inherently fluid, the resulting profiles are intended not as rigid classifications, but as heuristic, comparative guides to support curriculum design and textbook selection. Future work should examine inter-rater reproducibility across borderline cases, incorporate fine-grained indicators (such as prerequisites, exercise characteristics, and digital resources), explore additional entry-point configurations, including field-theoretic introductions to QM \cite{Hobson_2013, Giliberti_2024}, and, for the overall pedagogical orientation, distinguish mathematical orientation from rigor and technical sophistication, and experimental illustration from technological applications.

Finally, because textbooks represent intended curricula rather than enacted instruction, our findings motivate several directions for empirical PER. First, regarding \textit{mathematical tools and physical systems}, researchers should investigate how scaffolding advanced formalisms across instruction levels supports cross-representational fluency, and how integrating further contemporary experimental platforms---from solid-state and coupled spin-oscillator setups [Haroche '06$^1$; Simon '25$^1$] to neutron interferometry, quantum beats, and photon antibunching [Ballentine '15$^2$]---affects student sensemaking compared with idealized thought experiments. Second, regarding \textit{theoretical concepts}, fine-grained studies should examine how different formulations and contextualizations of postulates and principles, and particularly the classical--quantum relationship, shape students' ability to distinguish fundamental assumptions, derived results, and interpretational claims, as well as how formal instruction engages with students' prior popular conceptions of quantum phenomena (e.g., "Schrödinger’s cat" or "spooky action at a distance"). Third, regarding \textit{pedagogical pathways}, classroom studies are needed to test how concurrent integration and sequential combinations of entry points function for students with diverse preparation, and how alignment or discontinuity between introductory and advanced courses shapes conceptual understanding, problem-solving, and transfer across the undergraduate-to-graduate trajectory.

\section*{Acknowledgements}
We gratefully acknowledge discussions with Daniel Laumann. M.P. acknowledges support from EIN Quantum NRW.

\bibliography{references.bib, textbooks.bib}

\clearpage

\onecolumngrid 

\begin{center}
  \textbf{\large Supplementary Material}\\[.2cm]
\end{center}
\vspace{1cm}

\setcounter{equation}{0}
\setcounter{figure}{0}
\setcounter{table}{0}
\setcounter{page}{1}
\setcounter{section}{0}

\setcounter{figure}{0}
\renewcommand{\thefigure}{S\arabic{figure}}
\setcounter{table}{0}
\renewcommand{\thetable}{S\arabic{table}}
\setcounter{section}{0}
\renewcommand{\thesection}{S\arabic{section}}

\section{Codebook for RQ1}
\label{sec:codebook}
\paragraph{General coding rule.}
Unless a topic-specific definition states otherwise, record the earliest location at which the text defines, explains, mathematically formulates, or applies the topic sufficiently to establish its meaning or role. A name, historical reference, forward reference, or unexplained appearance in an equation alone does not constitute a substantive introduction. 
\\
Annotations to chapter codes indicate features of the topic’s presentation at the recorded location, such as whether specific terminology is explicitly used. 

\subsection{Mathematical formalisms}

\begin{table}[H]
\renewcommand{\arraystretch}{1.2}
\setlength{\tabcolsep}{5pt}
\footnotesize
\resizebox{1.0\textwidth}{!}{
\begin{tabular}{p{0.16\textwidth}p{0.51\textwidth}p{0.23\textwidth}}
\toprule
\textbf{Topic} & \textbf{Criteria for coding} & \textbf{Example  \newline [Auletta '09$^1$]}\\
\midrule
Wave \newline mechanics & The wave function is introduced as a state representation and used explanatory or computationally. If \textit{wave mechanics} does not appear, the chapter number is enclosed in parentheses. & 2/17 \newline  [pp. 39, pp. 68]  \\
Matrix \newline  mechanics & States and operators are represented by basis coefficients or matrix elements in a discrete basis, and matrix operations are used to describe or analyze a physical system. Finite- and infinite-dimensional representations both qualify. If \textit{matrix mechanics} does not appear, the chapter number is enclosed in parentheses.  & (1), 2/17 \newline  [pp. 20-28, pp. 39, pp. 48] \\
Dirac notation & Bra-ket notation is explicitly used for states, operators, or changes of representation. & 1/17 \newline  [pp. 22]  \\
Path integral & A \textit{path integral} is defined. &  10/17 \newline  [pp. 389-398] \\
Density matrix & A \textit{density matrix} or \textit{density operator} is defined. &  5, 9, 14/17 \newline  [pp. 174-190, pp. 278-350, pp. 513-529] \\
Second \newline quantization & \textit{Second quantization} is introduced; alternatively, creation and annihilation operators, occupation-number states, or Fock space are used to describe variable particle numbers or multiple identical degrees of freedom. If \textit{second quantization} does not appear, the chapter number is enclosed in parentheses. The treatment of ladder operators for a single harmonic oscillator is excluded unless they are connected to a many-particle occupation-number or quantum-field description. &  13/17 \newline  [pp. 457-506] \\
\bottomrule
\end{tabular}}
\end{table}

\begin{itemize}

\item If at least three mathematical formalisms are introduced in the same chapter as the first formalisms, the subchapter number is also noted to indicate the order in which they first appear.

\item Repeated discussions of path integrals, density matrices, and second quantization are also recorded when these topics receive detailed treatment in more than one section.

\end{itemize}


\subsection{First physical systems}

Conceptual illustrations (concerning single-photon systems) are distinguished from worked examples. A conceptual illustration is a physical system used primarily to motivate or interpret a quantum idea, without sustained mathematical treatment. A worked example is a treatment of a specified physical system in which the text performs a calculation or a step-by-step derivation leading to a mathematical result with a physical interpretation, such as eigenstate determination, probability computation, expectation values, or time evolution. The state may be given, constructed, or determined within the example. 

\begin{table}[H]
\renewcommand{\arraystretch}{1.2}
\setlength{\tabcolsep}{5pt}
\footnotesize
\resizebox{1.0\textwidth}{!}{
\begin{tabular}{p{0.23\textwidth}p{0.48\textwidth}p{0.21\textwidth}}
\toprule
\textbf{Topic} & \textbf{Criteria for coding}  & \textbf{Example  \newline [Auletta '09$^1$]} \\
\midrule
First worked example (infinite-dim. )  & The first worked example where the state space of the physical system used in the example’s model is infinite-dimensional.  &  infinite square well; 3/17 \newline  [pp. 113-116]\\
First worked example (finite-dim. )  & The first worked example where the state space of the physical system used in the example’s model is finite-dimensional.  &  Mach-Zehnder interferometer; 1/17 \newline  [pp. 14-16] \\
Qubit & The term \textit{qubit} is explicitly introduced or used with explanatory intent. &  17/17 \newline  [pp. 642-643] \\
Single-photon system & A single-photon system is used to explain or model quantum concepts such as superposition, interference, measurement, entanglement, nonlocality, or quantum information, rather than being mentioned only in historical discussions of early quantum theory. &  Mach-Zehnder interferometer; 1/17 \newline  [pp. 14-16]\\
\bottomrule
\end{tabular}}
\end{table}

\begin{itemize}
\item If more than two systems are introduced in the same chapter as first examples, the subchapter number is also noted to indicate the order in which they first appear.
\end{itemize}

\subsection{Postulates, principles, foundations and interpretations}

\begin{itemize}
    \item When a topic was used without an identifiable introduction, record it as \textit{Assumed}. This designation describes its treatment in the text and does not require an explicit prerequisite statement by the author.
    \item Repeated discussions of a topic are also recorded when it receives detailed treatment in more than one section. 
\end{itemize}

\subsubsection{Foundational postulates of QM}
A topic is recorded even if the exact term is absent or the exact designation (\textit{postulate}, \textit{rule}, \textit{principle}, or \textit{theorem}) does not appear; in such cases, the chapter number is enclosed in parentheses.

\begin{table}[H]
\renewcommand{\arraystretch}{1.2}
\setlength{\tabcolsep}{5pt}
\footnotesize
\resizebox{1.0\textwidth}{!}{
\begin{tabular}{p{0.19\textwidth}p{0.50\textwidth}p{0.20\textwidth}}
\toprule
\textbf{Topic} & \textbf{Criteria for coding} & \textbf{Example  \newline [Auletta '09$^1$]} \\
\midrule
State & Quantum states are associated with Hilbert space. &  (1)/17 \newline [pp. 25]\\
Observable & Physical observables are associated with operators. &  2$^\dagger$/17 \newline  [pp. 44]\\
Measurement \newline outcome &  Eigenvalues are identified as the possible outcomes of a measurement. Alternatively, expectation values are associated with measurement outcomes. &  2$^\dagger$/17 \newline  [pp. 57]\\
Measurement probability & A rule connecting the state to outcome probabilities, such as squared amplitudes or an equivalent projector/trace formulation. Annotate the chapter code to indicate when \textit{Born’s rule} or \textit{Born’s interpretation} (instead of \textit{postulate}) is explicitly used. &  ([1]), 2$^\dagger$/17 \newline  [pp. 28-31, pp. 57] \\ 
State after \newline measurement & An explanation or rule for the post-measurement state conditional on a specified measurement outcome. & (1)/17  \newline  [pp. 22] \\
Time evolution & The time-dependent Schrödinger equation is introduced. Annotate the chapter code to indicate when unitary time evolution is identified as the fundamental rule of dynamics. & \{(3)\}/17 \newline  [pp. 101-102] \\
Composite system &  Joint-state structure is defined or described, such as a tensor-product construction or equivalent treatment of joint states.  &  (6)/17 \newline  [pp. 228]\\
Symmetrization & The symmetrization requirement for identical particles is introduced. Annotate the chapter code to indicate when the \textit{spin–statistics theorem} (instead of the symmetrization postulate) is introduced.  &  [7]/17 \newline  [pp. 249]\\
\bottomrule
\end{tabular}}
\footnotesize
\textit{Note}: Symbols used in the example column are presentation conventions and are explained in the corresponding data table.

\end{table}

\subsubsection{Foundational principles/theorems of QM}
A topic is recorded even if the exact term is absent or the exact designation (\textit{postulate}, \textit{rule}, \textit{principle}, or \textit{theorem}) does not appear; in such cases, the chapter number is enclosed in parentheses.

\begin{table}[H]
\renewcommand{\arraystretch}{1.2}
\setlength{\tabcolsep}{5pt}
\footnotesize
\resizebox{1.0\textwidth}{!}{
\begin{tabular}{p{0.21\textwidth}p{0.50\textwidth}p{0.18\textwidth}}
\toprule
\textbf{Topic} & \textbf{Criteria for coding} & \textbf{Example \newline  [Auletta '09$^1$]} \\
\midrule
Superposition & Linear combinations of quantum states are introduced. &  1/17 \newline  [pp. 18]\\
Complementarity & Complementarity is introduced.  Annotate the chapter code to indicate when only wave-particle duality is discussed without broader complementarity framing.&  1/17 \newline  [pp. 19]\\
Uncertainty & An uncertainty or indeterminacy relation is introduced. Annotate the chapter code to indicate when \textit{Heisenberg indeterminacy relation}, \textit{Uncertainty relation}, or \textit{Heisenberg inequality} is explicitly used. &  2/17 \newline  [pp. 82-92]\\
Correspondence & A correspondence rule connecting quantum and classical descriptions is introduced. &  2/17 \newline  [pp. 72]\\
Exclusion & Pauli's exclusion principle is introduced. &  7/17 \newline  [pp. 251]\\
No-cloning theorem & The \textit{no-cloning} feature of quantum states is introduced. &  15/17 \newline  [pp. 548]\\
\bottomrule
\end{tabular}}
\end{table}

\subsubsection{Selected topics in foundations and interpretations}
A topic is recorded even if the exact term does not appear; in such cases, the chapter number is enclosed in parentheses.

\begin{table}[H]
\renewcommand{\arraystretch}{1.2}
\setlength{\tabcolsep}{5pt}
\footnotesize
\resizebox{1.0\textwidth}{!}{
\begin{tabular}{p{0.16\textwidth}p{0.50\textwidth}p{0.18\textwidth}}
\toprule
\textbf{Topic} & \textbf{Criteria for coding} & \textbf{Example \newline [Auletta '09$^1$]} \\
\midrule
Entanglement & Entanglement or entangled states are introduced. &  16, 17/17 \newline  [pp. 567-673]\\
Nonlocality & Nonlocal feature of QM is introduced. &  16/17 \newline  [pp. 587]\\ 
Interpretations \newline of QM & Interpretational theories of QM, such as many-worlds, de Broglie-Bohm theory, QBism, are introduced.  Annotate the chapter code to indicate when the discussion is limited to the Copenhagen interpretation where \textit{Copenhagen interpretation} is explicitly mentioned. &  9/17 \newline  [pp. 284-291]\\
Measurement \newline problem & \textit{The measurement problem} is introduced.  Annotate the chapter code to indicate when the treatment is limited to Schrödinger's cat paradox. &  9/17 \newline  [pp. 278]\\
Decoherence & \textit{Decoherence} is introduced. &  9/17 \newline  [pp. 288-308]\\
\bottomrule
\end{tabular}}
\end{table}

\end{document}